\documentclass[iop,apj,numberedappendix,onecolumn]{emulateapj}
\slugcomment{{\sc Accepted to ApJ}}
\usepackage{natbib}
\usepackage{graphicx}
\usepackage{amsmath,amsfonts}
\usepackage{multirow}
\usepackage[parfill]{parskip}

\setcitestyle{notesep={; }}

\def\sgra{Sgr~A$^{\ast}$}
\def\it{\tilde{I}}

\def\pt{\tilde{P}}
\def\mb{\breve{m}}

\defcitealias{TMS}{TMS}

\begin{document}
\title{High Resolution Linear Polarimetric Imaging for the Event Horizon Telescope}
\shorttitle{}

\author{Andrew A. Chael\altaffilmark{1}, 
Michael D.\ Johnson\altaffilmark{1}, 
Ramesh Narayan\altaffilmark{1}, 
Sheperd S.\ Doeleman\altaffilmark{1,2},
John F. C. Wardle\altaffilmark{3},
Katherine L. Bouman\altaffilmark{4}}
\shortauthors{Chael et al.}
\altaffiltext{1}{Harvard-Smithsonian Center for Astrophysics, 60 Garden Street, Cambridge, MA 02138, USA}
\altaffiltext{2}{Massachusetts Institute of Technology, Haystack Observatory, Route 40, Westford, MA 01886, USA}
\altaffiltext{3}{Brandeis University, Physics Department, Waltham, MA 02454, USA}
\altaffiltext{4}{Massachusetts Institute of Technology, Computer Science and Artificial Intelligence Laboratory, 32 Vassar Street, Cambridge, MA 02139, USA}
\email{achael@cfa.harvard.edu}

\begin{abstract}
Images of the linear polarization of synchrotron radiation around Active Galactic Nuclei (AGN) identify their projected magnetic field lines and provide key data for understanding the physics of accretion and outflow from supermassive black holes. The highest resolution polarimetric images of AGN are produced with Very Long Baseline Interferometry (VLBI). Because VLBI incompletely samples the Fourier transform of the source image, any image reconstruction that fills in unmeasured spatial frequencies will not be unique and reconstruction algorithms are required. In this paper, we explore extensions of the Maximum Entropy Method (MEM) to linear polarimetric VLBI imaging. In contrast to previous work, our polarimetric MEM algorithm combines a Stokes $I$ imager that uses only bispectrum measurements that are immune to atmospheric phase corruption with a joint Stokes $Q$ and $U$ imager that operates on robust polarimetric ratios. We demonstrate the effectiveness of our technique on 7- and 3-mm wavelength quasar observations from the VLBA and simulated 1.3-mm Event Horizon Telescope observations of Sgr A* and M87. Consistent with past studies, we find that polarimetric MEM can produce superior resolution compared to the standard CLEAN algorithm when imaging smooth and compact source distributions. As an imaging framework, MEM is highly adaptable, allowing a range of constraints on polarization structure. Polarimetric MEM is thus an attractive choice for image reconstruction with the EHT. 
\end{abstract}


\section{Introduction}
Magnetic fields in plasmas around compact objects such as pulsars and black holes are critical in powering their energetic emission. In Active Galactic Nuclei (AGN), magnetic fields in accretion disks around central supermassive black holes regulate outflow and interaction of the AGN with its host galaxy \citep{Fabian_AGN, Kormandy_BH}. Stochastic magnetic fields in the disk drive turbulence via the magneto-rotational instability \citep{Bal_Hawley}, enabling efficient accretion and conversion of gravitational energy into near-Eddington luminosities. Near the horizon, magnetic fields can convert spin energy from the black hole into energetic jets of plasma \citep{BZ, Bl_Payne}. Synchrotron emission from electrons in AGN cores and jets is characteristically linearly polarized, with a polarization direction that traces magnetic field lines in the plasma \citep{Ginzburg_Sync}. Faraday rotation of the polarization vector as it propagates through the plasma is determined by the magnetic field strength and electron density along the line of sight \citep{RL}. Thus, measurements of the linear polarization direction and its frequency dependence characterize the magnetic field around these objects and provide an observational windows into their fundamental physics. For a recent review on observations of magnetic fields in AGN, see \citet{Wardle_jets}.

Linear Polarimetric Very Long Baseline Interferometry (VLBI) at centimeter and millimeter wavelengths can measure polarization magnitude and orientation at high angular resolutions of fractions of a milliarcsecond \citep{RWB_1994,Attridge_2001,Attridge_2005}. VLBI observations provide an incomplete sample of the Fourier transform of the sky flux density distribution, so image reconstruction (or deconvolution) algorithms are needed. The standard reconstruction algorithm is CLEAN \citep{Hog_1974}, which models the image as a collection of point sources. CLEAN produces images of the three Stokes parameters $I$, $Q$, and $U$ separately, so unphysical fractional polarizations $m=\sqrt{Q^2 + U^2}/I>1$ are possible (especially in regions of low total intensity $I$). In contrast, Bayesian regularization methods can naturally incorporate prior information on the image's spatial distribution and physical constraints such as $m \leq 1$. One such regularization method is the Maximum Entropy Method, (MEM), which finds the image most consistent with the data that maximizes an entropy function, analogous to the log of a prior probability distribution. MEM imaging algorithms have been in use for decades (e.g., \citet{GS_1978} and \citet{Cornwell_1985}), but because of early computational limitations, they are used infrequently compared to CLEAN. The theory behind polarimetric MEM was pioneered in several theoretical papers beginning in the 70s \citep{Ponsonby_1973, NN_1983, NN_1986, Shev_1987} but implementation on actual VLBI data has been limited to only a handful of studies since \citep{HW_MEM_1990, HoldThesis_1990, Sault_1999, Coughlan_2012, Coughlan_2013}.

Polarimetric MEM is particularly promising for imaging the accretion flow and jets near supermassive black holes observed by the Event Horizon Telescope (EHT). The EHT is a global 1.3-mm VLBI array that will obtain nominal resolutions of approximately 25 microarcseconds, allowing observations of nearby supermassive black holes at scales on the order of the projected Schwarzschild radius \citep{Decadal}. Previous observations with three EHT baselines have constrained the size of the 1.3-mm emission region to scales on order of the lensed Schwarzschild radius in \sgra\ \citep{Doeleman_2008, Fish_2011} and M87 \citep{Doeleman_2012}. First polarimetric observations with the EHT in 2013 resolved the polarimetric emission in \sgra\ providing strong evidence for ordered magnetic fields near the event horizon \citep{Johnson_Science}. Future observations are expected to obtain enough data to construct an image of the \sgra\ black hole accretion flow \citep{Fish_2014} and jet base of M87 \citep{Lu_2014}. As the EHT is now capable of observing with full polarization, polarimetric MEM provides an attractive solution to creating full-polarization images of these sources.

In this paper, we develop an application of the Maximum Entropy Method to polarimetric VLBI data. In Section~\ref{sec::imaging}, we review the fundamentals of polarimetric VLBI, as well as the standard CLEAN algorithm for image reconstruction. In Section~\ref{sec::IMEM}, we review standard MEM applied to total intensity, or Stokes $I$, images. We discuss applications of MEM to data without calibrated phase information, and quantify the resolution and fidelity of MEM images. We then move to polarimetric MEM in Section~\ref{sec::PolMEM}, where we introduce and compare two forms of the polarimetric entropy function. 

In Section~\ref{sec::results}, we discuss our implementation of polarimetric MEM, including the details of our simulation of EHT data and our minimization algorithm. We present results from applying  polarimetric MEM to a 7-mm VLBA observations of the quasar 3C279 and 3-mm VLBA observations of 3C273, as well as simulated EHT data from several 1.3-mm model images of \sgra\ and M87. We compare the effects of different forms of the entropy function, and we test the ability of polarimetric MEM to resolve polarization field structure. We show that for these high-frequency VLBI observations, polarimetric MEM is capable of reproducing the general morphology of the CLEAN images, but with typically higher resolution and fidelity. Finally, in Section~\ref{sec::Summary}, we outline future directions for polarimetric MEM in VLBI and synthesis imaging in interferometry more broadly. 

Our imaging code is written in Python and uses the L-BFGS minimization routine in the Scipy package. Our imaging programs as well as a variety of routines for simulating and manipulating VLBI data are available at \url{https://github.com/achael/eht-imaging}.

\section{Fundamentals of Interferometric Imaging}
\label{sec::imaging}

By the Van Cittert-Zernike theorem, measured interferometric visibilities $\tilde{I}_k$ are the Fourier components of the true source image $I(x,y)$ (in total intensity) plus thermal noise $n_k$ \citep{TMS}(hereafter TMS):
\begin{align}
 \label{eq::VCZ}
 \it_k=\int\int I(x,y)e^{-2\pi i (u_kx+v_ky)}\mathrm{d}x\,\mathrm{d}y + n_k.
\end{align}
Here, $x$ and $y$ are real space angular coordinates and $u_k$, $v_k$ are the interferometric baseline coordinates projected orthogonal to the source line of sight and measured in wavelengths. 

An interferometer incompletely samples the $u-v$ plane, so direct Fourier transform (i.e., the ``dirty image'') of the measured visibilities is a convolution of the true image and the Fourier transform of the $u-v$ sample coverage (the ``dirty beam''). VLBI imaging can thus be approached either as deconvolution of the dirty beam from the dirty image or as fitting a model to visibility data with regularizing constraints. Finite sampling also ensures that no image that reproduces the observed visibilities $\tilde{I}_k$ is unique; extra information is always required to constrain the image. In CLEAN, this extra information is the representation of the sky image in terms of a finite number of point sources. MEM allows for many potential regularizing constraints through the use of different entropy functions. Furthermore, MEM naturally incorporates uncertainties due to thermal noise and quantifies the goodness-of-fit in a standard $\chi^2$ metric. This makes MEM a natural choice for sparse or heterogeneous VLBI arrays such as the EHT.

A further complication in total-intensity VLBI at high frequencies is that atmospheric fluctuations make stable phase information on individual baselines impossible. However, adding the phases observed on three baselines around a triangle cancels the atmospheric contribution at each station, so these \emph{closure phases} contain only information about the source. Assuming the visibility amplitudes can be calibrated to remove station-dependent gain terms (which vary more slowly than the unstable phase terms from the atmosphere), calibrated amplitudes can be combined with closure phases in the bispectrum, the product of three simultaneous visibilities around a triangle  \citep{Rogers_74},\citepalias{TMS},
\begin{equation}
 \label{eq::bispec}
 \it_B = \it_{12}\,\it_{23}\,\it_{31},
\end{equation}
where, for instance, $\it_{12}$ is the measured visibility between stations $1$ and $2$ at a given time. 

Because the image of linear polarization is a two-dimensional vector field defining both the magnitude and direction of the linear polarization at each location, it can be represented as a complex image, $P(x,y) = Q(x,y) + iU(x,y)$, where $Q(x,y)$ and $U(x,y)$ are the images of the linear Stokes parameters. The linearly polarized image can also be expressed in terms of the polarization fraction $m(x,y)$ and polarization position angle $\chi(x,y)$ as $P = m\,I\,e^{2i\chi}$, where $m = |P|/I = \sqrt{Q^2+U^2}/I \leq 1$, and $\chi = \frac{1}{2}\arctan\frac{U}{Q}$. Polarimetric visibilities, $\tilde{Q}_k$ and $\tilde{U}_k$, are also related to the images $Q(x,y)$ and $U(x,y)$ via the van Cittert-Zernike theorem (Eq.~\ref{eq::VCZ}). For synthesis imaging, the most significant difference between polarization and total-flux is that the images of $Q$ and $U$ are not constrained to be positive and that the total polarization fraction in each pixel is constrained to be less than one.

Because the atmosphere is not significantly birefringent at the high frequencies we are considering \citepalias{TMS}, the atmospheric contribution to phase is identical for all of the visibilities $\tilde{I}_k$, $\tilde{Q}_k$, and $\tilde{U}_k$, so polarimetric ratios such as $\tilde{Q}_k/\tilde{I}_k$ provide the same immunity to station-based phase errors as closure phase \citep{RWB_1994}.  We define the visibility domain polarimetric ratio and phase (following the notation of \citet{Johnson_2014})
\begin{equation}
 \label{eq::polrat}
 \breve{m}_k = \frac{\tilde{P}_k}{\tilde{I}_k}.
\end{equation}
It is important to note that $\mb$ is \emph{not} the Fourier transform of the image plane polarization fraction $m$. In particular, $\mb$ is not conjugate-symmetric under the reversal of baselines $(u,v) \rightarrow (-u,-v)$.  It also is possible for the magnitude $|\mb|$ to exceed unity, if for example the total intensity visibility $\tilde{I}$ has a ``null'' at some baseline due to the presence of some sharp feature in the image \citep{Johnson_Science}. 

\section{Total Intensity MEM}
\label{sec::IMEM}

The standard CLEAN algorithm operates on the dirty image obtained from Fourier transforming the sparse interferometer data and treats the imaging process as a deconvolution of the dirty beam from the image \citep{Hog_1974}. CLEAN models the sky brightness distribution as a collection of point sources. It determines the locations and magnitudes of these point sources iteratively by finding the maximum intensity pixel of the dirty image, then subtracting the shifted and scaled ``dirty beam''. After a certain number of iterations, CLEAN convolves the point source model with a ``clean'' beam obtained by fitting a Gaussian to the central component of the dirty beam. The algorithm halts after the maximum brightness point in an image drops below some multiple of the residual RMS level, or after negative components start to be removed. Finally, the dirty image residuals are added to the restored image to include low-intensity diffuse brightness distributions that are poorly captured by the point source decomposition. Because CLEAN relies on absolute visibility phase information to perform the inverse Fourier Transform to the dirty image at each step of the algorithm, visibility phases corrupted by atmospheric phase fluctuations must be either calibrated or self-calibrated in a loop with multiple iterations of CLEAN \citepalias{TMS}. 

In contrast, MEM as described in this paper operates directly on the measured visibilities or robust quantities like closure phases or the bispectrum. In MEM and other Bayesian regularization imaging methods, an image is fitted to the  data by minimizing a weighted sum of $\chi^2$ and a regularizing function which incorporates prior information. With this approach, only the forward Fourier transform from the trial image to the visibility domain is used, and trial-image visibilities can be directly compared with measured, calibrated visibilities or other data products derived from the measured visibilities. Furthermore, for sparse VLBI arrays, we can avoid sampling errors introduced by transforming with a Fast Fourier Transform (FFT) and compute trial-image visibilities at the sampled baseline points with a discrete-time Fourier Transform (DTFT).\footnote{While the term ``discrete-time Fourier transform'' refers to time as the discretely sampled interval, in our case the transform is spatial and the discretely sampled interval is the image angular coordinate.}

In MEM, we maximize a regularizing function, or ``entropy'' of the image with respect to data constraints. In what follows, we denote all arrays of image pixels or visibilities in bold. For a $n^2$ pixel test image $\mathbf{I'}$, a prior/bias image $\mathbf{B}$, and an array of $N$ measured visibilities $\mathbf{\tilde{I}}$, we maximize the objective function \citep{NN_1986}
\begin{align}
 \label{eq::objfunc}
 J = S(\mathbf{I'},\mathbf{B}) - \alpha \left(\chi^2(\mathbf{I'})-1\right),
\end{align}
where $S(\mathbf{I'},\mathbf{B})$ is the chosen regularizer or entropy function and $\chi^2$ is the goodness-of-fit test statistic that compares the visibilities of the test image $\mathbf{I'}$ to the data. The mixing coefficient $\alpha$ controls the weighting between the regularizer (entropy) term and the data ($\chi^2$) term. Considering MEM as a form of constrained optimization, $\alpha$ plays the role of a Lagrange multiplier. In practice, it can be fixed, iterated manually, or be allowed to vary in the maximization process.

The goodness-of-fit $\chi^2$ term is defined in the visibility domain:
\begin{equation}
 \label{eq::chi2}
 \chi^2(\mathbf{I'})=\frac{1}{2N}\sum\limits_{k=1}^{N} \frac{1}{\sigma_k^2}|\it_k-\it'_k|^2,
\end{equation}
where $\sigma_k$ is the noise estimate on the $k$th $u,v$ point and the model visibilities $I'_k$ are the DTFT of the test image evaluated at the $k$th $u,v$ point. The factor of 2 in the denominator of Eq.~\ref{eq::chi2} is included because the variance $\sigma_k^2$ is taken to be the variance along either the real or imaginary axis; in fitting the data, we must fit the real and imaginary parts of the visibilities separately. Assuming that the visibilities are normally distributed, $\chi^2$ will possess a $\chi^2$ distribution, and a good fit where the trial visibilities agree with the measurements within error has $\chi^2 \approx 1$.

The standard entropy function motivated from information theory is \citep{Frieden_1972, GS_1978}:
\begin{align}
 \label{eq::Ssimple}
 S(\mathbf{I'},\mathbf{B}) = -\sum\limits_{i=1}^{n^2} I'_i\log\left(\frac{I'_i}{B_i}\right),
\end{align}
but many other entropy functions can be chosen, including  $S(\mathbf{I'}) = \sum \log(I'_i)$, $S(\mathbf{I'}) = \sum \sqrt{I'_i}$ \citep{NN_1986}, or the $\ell_1$ norm $S(\mathbf{I'}) = \sum|I'_i|$ \citep{Honma_2014}. In fact, it can be shown that for any convex function $S(\mathbf{I'},\mathbf{B})$ of the $I_i$, the reconstruction is guaranteed to converge \citep{NN_1986}.

To deal with phase uncertainty in MEM without needing to calibrate or self-calibrate the visibility phases, we can extend this technique to the image bispectrum, replacing the data term $\chi^2$ with its bispectral extension. In this method, unlike in a self-calibration loop as used with CLEAN, the visibility phases are not calibrated prior to imaging. The objective function becomes
\begin{equation}
 \label{eq::objfuncbs}
 J_B = S(\mathbf{I'},\mathbf{B}) - \alpha (\chi^2_B(\mathbf{I'}) - 1),
\end{equation}
where the bispectrum data term is
\begin{equation}
 \label{eq::chisqbi}
 \chi_B^2(\mathbf{I'}) = \frac{1}{2N_B}\sum\limits_{j=1}^{N_B} \frac{1}{\sigma_{B\,j}^2}|\it_{B\,j}-\it'_{B\,j}|^2.
\end{equation}
In the above equation, we have $N_B$ \emph{independent} bispectrum measurements $\it_{Bj}$ each with standard deviation $\sigma_{Bj}$. At any instant in time with detections on all baselines to $T$ sites, there are ${T \choose 3} = T!/\,3!\,(T-3)!$ triangles but only $(T-1)(T-2)/2$ independent bispectrum measurements \citepalias{TMS}. In our reconstructions, we constructed a set of independent bispectra at each time using the criterion that each triangle contain the station with the highest signal-to-noise ratio.

Using the bispectrum for MEM image reconstruction was pioneered in optical interferometry with the BSMEM gradient descent algorithm \citep{BSMEM_94, BSMEM_08}, which has been successfully used on simulated EHT observations \citep{Fish_2014, Lu_2014}. Recent developments using the bispectrum directly in image reconstruction include the CHIRP algorithm \citep{Katie_2015}, which uses a data-driven regularizing function based on features found in a library of sample images instead of a standard entropy term based on a single prior image.

Because the bispectrum may not include any significantly small triangles to tightly constrain the unresolved flux, it is particularly useful to add a total flux constraint to the objective function (Eq.~\ref{eq::objfunc}):
\begin{align}
 \label{eq::Fconstraint}
 J_B \rightarrow J_B + \gamma\left[\sum_{i=1}^{n^2}{I'_i}-F_{\text{obs}}\right]^2,
\end{align}
where $F_\text{obs}$ is the observed total flux density of the source and $\gamma$ is a hyperparameter that controls the relative weighting of the flux constraint compared to the $\chi^2_B$ and entropy term in Eq.~\ref{eq::objfuncbs}. In addition, since the bispectrum lacks overall phase information, the bispectrum data carry no information about the absolute image position and the final image centroid is arbitrary. Choosing image frame coordinates where $x=0, y=0$ corresponds to the center of the frame, we can add a center-of-mass constraint to the objective function:
\begin{align}
 \label{eq::COMconstraint}
 J_B \rightarrow J_B + \delta\left[\sum_{i=1}^{n^2}{I'_i}x_i + \sum{I'_i}y_i\right]^2,
\end{align}
where $\delta$ is another hyperparameter to control the weight of this constraint with respect to the data term, entropy, and any other constraints. Finally, imaging with the bispectrum carries an additional complication in that the bispectrum values are not necessarily described by a normal distribution, and thus the bispectrum $\chi_B^2$ statistic is not as straightforward to interpret as in the case where calibrated visibility phases are available. In the high signal-to-noise limit, however, the distribution of the bispectrum points approaches a Gaussian \citep{RDM_1995}.

We can implement a total intensity MEM algorithm using a quasi-Newton gradient descent method. The derivatives of $S$ and $\chi^2$ or $S_B$ and $\chi^2_B$ with respect to the pixel values $I_i$ can be computed analytically and evaluated at each step. The inverse Hessian can be approximated by neglecting the off-diagonal terms, as in the Cornwell and Evans algorithm \citep{Cornwell_1985}, or through numerical approximation as in the Broyden-Fletcher-Goldfarb-Shanno (BFGS) algorithm \citep{LBFGS}.

Because nonlinear methods like MEM and CLEAN input prior information into the imaging process, we might expect some degree of image ``superresolution,'' or the production of image features on scales less than the array nominal resolution $R_\text{min}=\lambda/b_\text{max}$, where $b_\text{max}$ is the length of the longest baseline in the VLBI array. The frequently quoted result that MEM has a superresolution factor of $1/4$ the nominal resolution is in fact \emph{not} based on the image prior, but only on the analyticity of the data \citep{NN_1986}. The derivation of this factor requires the unrealistic assumption of infinite signal-to-noise \citep{HoldThesis_1990}. In practice, the superresolution factor may be informed by both the analyticity of the data (degraded by noise) and the entropy function. Simple tests comparing blurred MEM to the model source distribution suggest that, in practice, total intensity MEM can achieve a superresolution factor of $1/3$ to $1/2$ the nominal resolution (see section~\ref{sec::cleancompare}).

\section{Linear Polarimetric MEM}
\label{sec::PolMEM}

To extend MEM to complex linear polarized images while avoiding atmospheric phase corruption, we maximize the objective function \citep{HoldThesis_1990}
\begin{align}
 \label{eq::objfuncpol}
 J_m = S_m(\mathbf{P'}) - \beta \left(\chi^2_m(\mathbf{I'},\mathbf{P'})-1\right).
\end{align}
Here $\mathbf{P'}$ is our trial image of the polarized flux and $\chi^2_m$ is the test statistic that compares the polarimetric ratios of the test image to the data:
\begin{equation}
\label{eq::chi2m}
\chi^2_m(\mathbf{I'},\mathbf{P'}) = \frac{1}{2N}\sum\limits_{k=1}^{N} \frac{1}{\sigma_{m\,k}^2}|\mb_k-\mb'_k|^2,
\end{equation}
where $\mb_k = \pt_k/\it_k$ is the polarimetric ratio on the $k$th $u,v$ point, which is insensitive to phase errors. As in the bispectral imaging of total intensity, this MEM technique does not reconstruct the phases on $\tilde{Q}$ and $\tilde{U}$ directly before imaging, but instead relies on the robust, measurable polarimetric ratios $\breve{m}$ to guide the imaging directly. Thus, computing $\chi^2_m$ requires a $\mathbf{I'}$ reconstruction and its visibility phases - when using the polarimetric ratios in this manner, we cannot image $\mathbf{P'}$ independently from $\mathbf{I'}$.

The conventional polarimetric entropy, first developed from the eigenvalues of the Stokes parameter correlation matrix, is of the form \citep{Ponsonby_1973, NN_1983, NN_1986, HW_MEM_1990}
\begin{align}
 \label{eq::HWent}
 S_m(\mathbf{P'}) = -\sum\limits_{i=1}^{n^2}I'_i\left[\frac{m_{\text{max}}+m'_i}{2}\log\left(\frac{m_{\text{max}}+m'_i}{2}\right)+ \frac{m_{\text{max}}-m'_i}{2}\log\left(\frac{m_{\text{max}}-m'_i}{2} \right)\right].
\end{align}
The quantity $m_\text{max}$ is the maximum fractional polarization; this can be generically set to 1. For synchrotron sources we can instead set $m_\text{max} \approx 0.75$, as was done by \citet{HW_MEM_1990} to ensure that the degree of polarization remains limited to the expected maximum for power-law synchrotron emission \citep{RL}. This entropy naturally favors images with $|m'_i|<m_\text{max}$. However, it contains no information about the polarization direction and it tends to favor low polarization magnitudes, as it is maximized in the absence of data constraints when $m'_i = 0$ in all pixels.

Another possibility is to use a simple log entropy as one might use for total intensity images (e.g., \citet{Ponsonby_1973,NN_1983,NN_1986}):
\begin{align}
 \label{eq::Ssimplepol}
 S_m(\mathbf{P'}) = -\sum\limits_{i=1}^{n^2} |P'_i|\log |P'_i|.
\end{align}

Because the polarization field traces the magnetic field structure, which we expect to not be completely disordered in resolved images, we may want to move beyond pixel-by-pixel entropy terms to gradient-based regularizing functions. One option is to use a regularizer proportional to the total variation of the trial image $\mathbf{P'}$. The isotropic total variation $TV$ of a  complex image matrix $\mathbf{X}$ typically used for image reconstruction and denoising \citep{TV} is:
\begin{align}
 \label{eq::tv}
 TV(\mathbf{X}) = \sum\limits_{i=1}^{n}\sum\limits_{j=1}^n \sqrt{|X_{i+1,j}-X_{i,j}|^2 + |X_{i,j+1}-X_{i,j}|^2}.
\end{align}
Adopting the total variation of the complex polarimetric image as our ``entropy'' term, we take $S(\mathbf{P'}) = -TV(\mathbf{P'})$.

The total variation entropy constrains MEM to prefer smooth polarization fields in both direction and magnitude. However, the gradient of Eq.~\ref{eq::tv} becomes infinite for uniform images, so care must be taken in the minimization algorithm, especially in determining the initial test image. The actual imaging algorithm can again use quasi-Newton or conjugate gradient methods, and can operate on either the $Q$ and $U$ arrays or the $m$ and $\chi$ images.

Finally, we can combine the polarimetric and total flux terms into a joint objective function for simultaneous imaging of $\mathbf{I'}$ and $\mathbf{P'}$ with the bispectrum and polarimetric ratios:
\begin{align}
 \label{eq::objfuncjoint}
 \begin{split}
 J_\text{tot} = & S_\text{tot}(\mathbf{I'},\mathbf{P'},\mathbf{B}) - \alpha (\chi^2_B(\mathbf{I'}) - 1) - \beta (\chi^2_m(\mathbf{I'},\mathbf{P'}) - 1) + \text{constraints}.
 \end{split}
\end{align}
Here $S_\text{tot}(\mathbf{I'},\mathbf{P'},\mathbf{B})$ is a joint entropy function, such as a combination of Eq.~\ref{eq::Ssimple} and Eq.~\ref{eq::HWent}, and the constraints can include terms constraining the total flux density (Eq.~\ref{eq::Fconstraint}) or image centroid (Eq.~\ref{eq::COMconstraint}).  Joint imaging can use data in the polarimetric ratios to constrain the total intensity image, and is thus the most theoretically sound method of MEM imaging polarized fields. In practice, however, convergence to the true image is poor if we allow both $\mathbf{I'}$ and $\mathbf{P'}$  to vary starting from a flat or Gaussian initial image \citep{HW_MEM_1990}.  The method favored by \citet{HoldThesis_1990} alternates iterations where $\mathbf{I'}$ and $\mathbf{P'}$ are changed independently. In our experience, this method does not offer any practical benefit over imaging $\mathbf{I'}$ and $\mathbf{P'}$ separately, as the $\mathbf{P'}$ reconstruction is not allowed to constrain the $\mathbf{I'}$ image. To aid convergence, we adopted a strategy of performing initial imaging steps with total intensity and polarization separately and then using the result as the initial image of a joint imaging process. Even in this case, we found the performance of the joint imaging routine to be highly sensitive to the weighting between the bispectrum and polarimetric ratio data terms ($\alpha$ and $\beta$ in Eq.~\ref{eq::objfuncjoint}), and the advantage over careful independent imaging of $\mathbf{I'}$ and $\mathbf{P'}$ seems minimal.

\section{Implementation and Results}
\label{sec::results}
In this section, we discuss the implementation of our MEM algorithm and results from applying it to real and simulated data sets. In section~\ref{sec::imp}, we describe the algorithm's implementation in our python software (available at \url{https://github.com/achael/eht-imaging}). In section~\ref{sec::bu}, we check our method for consistency with CLEAN on established reconstructions of VLBA observations of the quasar 3C279 at 7-mm and of 3C273 at 3-mm. In section~\ref{sec::cleancompare}, we characterize our algorithm's ability to ``superresolve'' image structure and compare it's performance with CLEAN's in the regime of model images of \sgra\ at 1.3-mm, as might be observed by the EHT. Finally, in section~\ref{sec::EHTobs}, we apply our method to several different simulated EHT data sets from the array that observed \sgra\ and M87 in March 2016 and the expected expanded array in 2017 and interpret the results.

\subsection{Implementation}
\label{sec::imp}
We imaged a variety of simulated and real polarimetric VLBI data sets by numerically maximizing the polarimetric ratio objective function, Eq.~\ref{eq::objfuncpol}, given Stokes $I$ images generated by MEM on the bispectrum obtained by minimizing Eq.~\ref{eq::objfunc}. Because of the relatively small size of our VLBI data sets and the limited fields of view of our reconstructed images, we used DTFTs instead of FFTs in calculating the $\chi^2$ terms and the gradients of $J_B$ and $J_m$. This eliminates the error introduced by interpolating data from the measured $u,v$ points to the FFT grid. Our MEM routines use the Limited-Memory BFGS algorithm (L-BFGS) \citep{LBFGS} implemented in the Scipy scientific python package \citep{LBFGS2, SCIPY}. L-BFGS is a quasi-Newton gradient descent algorithm which relies on a gradient function provided by the user and estimates the Hessian matrix as it iterates. L-BFGS does not store a full Hessian matrix, but approximates it with a series of vectors from the preceding $m$ steps, making it a preferred choice for our reconstructions due to the large size of the Hessian ($n^2 \times n^2$ for an $n \times n$ image). On our limited data sets, we found that L-BFGS ran sufficiently quickly on images up to $500 \times 500$ pixels, and it was consistently more accurate in its reconstructions of model images than either a polarimetric modification of the Cornwell-Evans algorithm or a conjugate gradient method.

A major difference between our MEM implementation and CLEAN is that MEM uses only a forward transform from the sample image to visibility space while CLEAN uses inverse transforms from the visibilities to the image domain. In CLEAN, the inverse transforms require the visibility data to be gridded, but in MEM the sample visibilities can naturally be computed with a DTFT using the exact sampled $u$-$v$ points.  Our method is also different than some past MEM algorithms (e.g. \citet{Cornwell_1985}) which have generally used the forward transform, approximating the data $\chi^2$ as a difference between the dirty image and the test image convolved with the dirty beam. Our method instead uses DTFTs and is ideal for sparse VLBI arrays. 

While entropy terms like Eq.~\ref{eq::Ssimple} and Eq.~\ref{eq::HWent} can be designed to prefer images that obey physical constraints like $I > 0$ and $0 < m < 1$, the initial steps of an unbounded minimization algorithm often take the image into an unphysical configuration and complex values of the entropy functions. This problem can be addressed by using a bounded minimization algorithm or placing a manual clip on the values of $I$ or $m$ (as in \citet{Cornwell_1985}). Instead, we chose to perform a change of variables in both the total intensity and polarimetric images to naturally enforce the image constraints.

In Stokes $I$, to satisfy the total intensity constraint $I \geq 0$, we transformed to $I=e^\xi$, where $-\infty<\xi<\infty$. For the polarimetric data, we choose to reconstruct images in $m$ and $\chi$, instead of $Q$ and $U$, as both the physical constraint $m<1$ and the entropy functions Eq.~\ref{eq::HWent} and Eq.~\ref{eq::tv} are most naturally defined in terms of these variables.  To naturally satisfy the constraint $0<m<1$, we transformed to $m=\frac{1}{2}+\frac{1}{\pi}\arctan \kappa$, where $-\infty<\kappa<\infty$. In both cases, we modified the gradient given to the algorithm by multiplying by the derivatives of these expressions (See Appendix~\ref{sec::cov}).

\begin{figure*}[h!]
\centering
\includegraphics*[width=.65\linewidth]{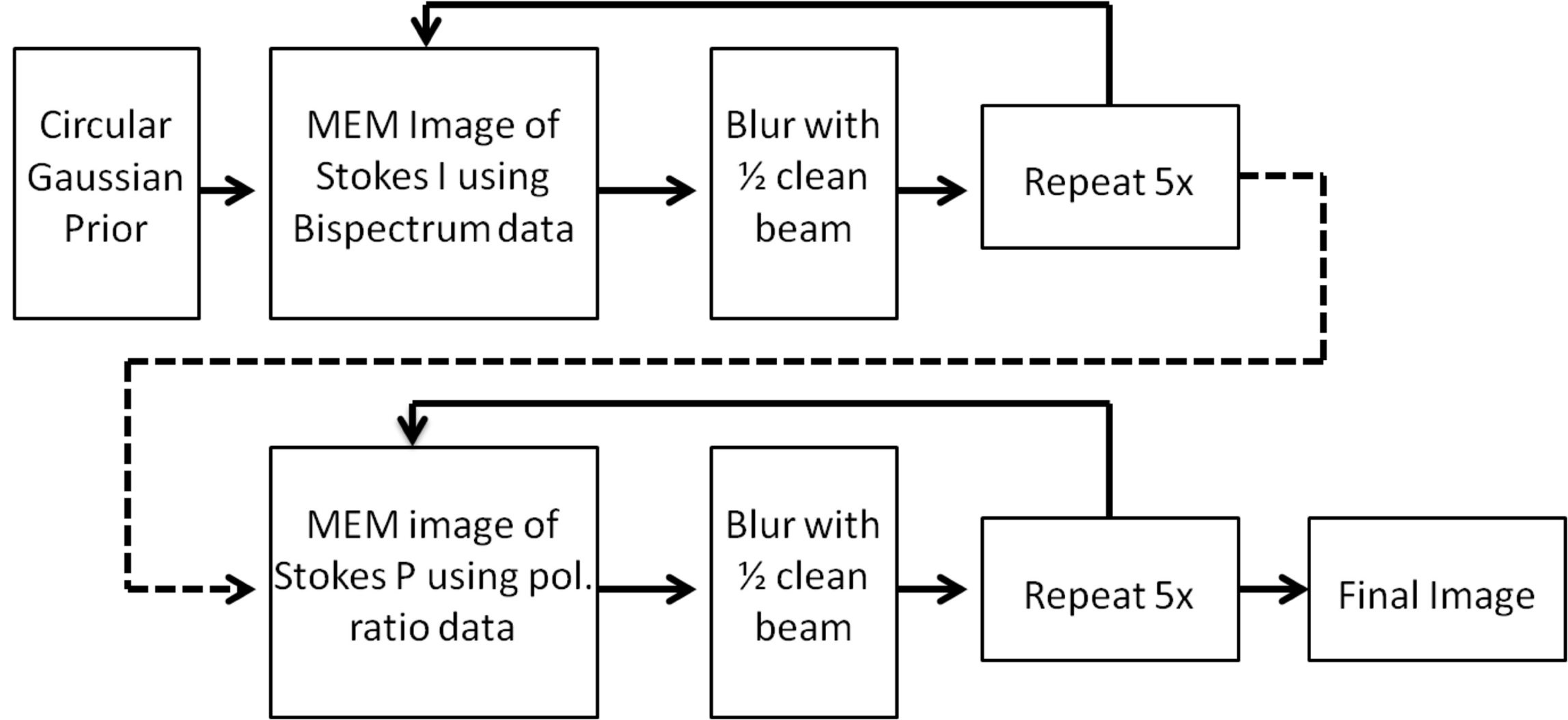}
\caption
{
Flowchart summarizing our imaging procedure.
}
\label{fig::proc}
\end{figure*}

To compare polarimetric regularizers, we used both the standard entropy term, Eq.~\ref{eq::HWent}, which we refer to as the Ponsonby-Nityananda-Narayan (PNN) entropy, and a total variation entropy term, Eq.~\ref{eq::tv}. In all of our reconstructions, we first imaged $\mathbf{I'}$ directly using the bispectrum. Because bispectral imaging of $\mathbf{I'}$ can converge poorly given a poor choice of image prior, we used a sequence of five runs of the algorithm substituting the prior image in Eq.~\ref{eq::Ssimple} with the final image from previous run blurred with a 1/2 scaled clean beam. To initialize the $P$ imaging process, we use an initial image that has constant fractional polarization magnitude and direction, set equal to the zero-baseline value, which is multiplied by the final $\mathbf{I'}$ image.  Our tests have shown, however, that the final polarimetric image is generally insensitive to the initial polarimetric image. If zero baseline polarization data is not available, an image with constant 5\% polarization fraction and zero polarization position angle may be used instead. We again used a sequence of five runs of imaging $\mathbf{P'}$, using the final image blurred to 1/2 the array resolution as the initial image of each subsequent run. Finally, we again convolved the final $\mathbf{I'}$ and $\mathbf{P'}$ images with a Gaussian beam 1/2 the size the of the fitted clean beam, limiting MEM's tendency to superresolve spurious features. Note that since the bispectrum does not contain absolute phase information constraining the location of the total flux image centroid, MEM images produced with the bispectrum are frequently offset from the model image, despite attempts to constrain this tendency with Eq.~\ref{eq::COMconstraint}. As a result, we have manually centered our images to provide clear comparisons with the model images. The essential steps of our procedure are summarized in Fig.~\ref{fig::proc}.

Our python code with routines for simulating and manipulating data and producing MEM images with the Scipy L-BFGS algorithm is available to download at \url{https://github.com/achael/eht-imaging}.

\subsection{7-mm and 3-mm VLBA quasar observations}
\label{sec::bu}

\begin{figure*}[h!]
\centering
\includegraphics*[width=0.8\linewidth]{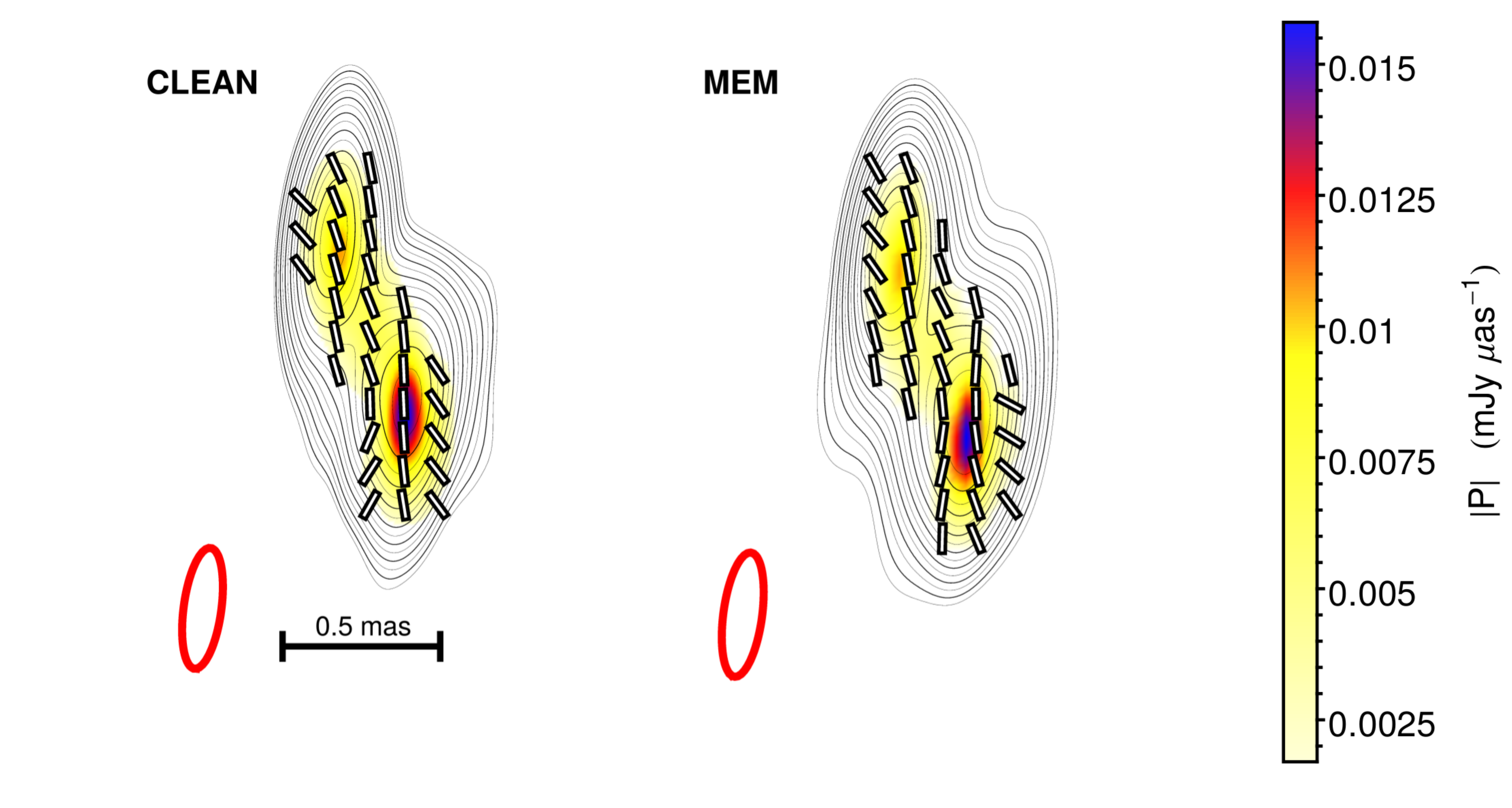}
\caption
{
Reconstructions of 7-mm observations of the quasar 3C279 taken with the VLBA in April 2013 \citep{Svetlana}. Contours are of total flux in steps of $\sqrt{2}$ from 3$\times$ the background RMS level. Ticks representing the direction of the polarization position angle and color corresponding to the polarized intensity $|P|$ are plotted in regions where $|P|$ is greater than 4$\times$ its background RMS level. The left panel shows the reconstruction convolved with the fitted elliptical clean beam ($384 \times 119$ $\mu$as FWHM, produced with Briggs weighting), and the right panel displays a MEM reconstruction of the same data set, smoothed with the fitted clean beam. While the CLEAN reconstruction used a self-calibration loop to determine visibility phases on $\tilde{I}$, $\tilde{Q}$, and $\tilde{U}$, the MEM reconstruction directly used bispectrum and polarimetric ratio data. The MEM reconstruction used the Ponsonby-Nityananda-Narayan (PNN) entropy term. The results are consistent with the CLEAN reconstruction when convolved with the same beam. 
}
\label{fig::BUapril}
\end{figure*}

To check consistency with CLEAN on real data, we produced polarimetric images from 7~-mm quasar observations from the Very Long Baseline Array obtained by the Boston University Blazar Research Group in 2013 (data reduction is described in \citet{Svetlana}).\footnote{http://www.bu.edu/blazars/VLBAproject.html} We compared our polarimetric images with both established CLEAN reconstructions convolved with the fitted clean beam and high-resolution versions of the reconstruction convolved with a circular $0.1\times0.1$ milliarcsecond beam.  While the VLBA data were phase-calibrated, we still used MEM algorithms that only used bispectrum and polarimetric ratio data. Our results for the quasar 3C279 are displayed in Fig.~\ref{fig::BUapril}. We found when our MEM reconstructions were convolved with the same beam used by CLEAN, the MEM images were an excellent match for the overall polarization magnitude and direction structure of the established images. Both the PNN (Eq.~\ref{eq::HWent}) and total variation regularizers (Eq.~\ref{eq::tv}) performed well in reconstructing the direction of the polarization field, and were consistent with each other. In general, the total variation regularizer, which does not prefer low polarization magnitudes, produced higher fractional polarization than the PNN regularizer in areas with weak Stokes I flux.

\begin{figure*}[h!]
\centering
\includegraphics*[width=1\linewidth]{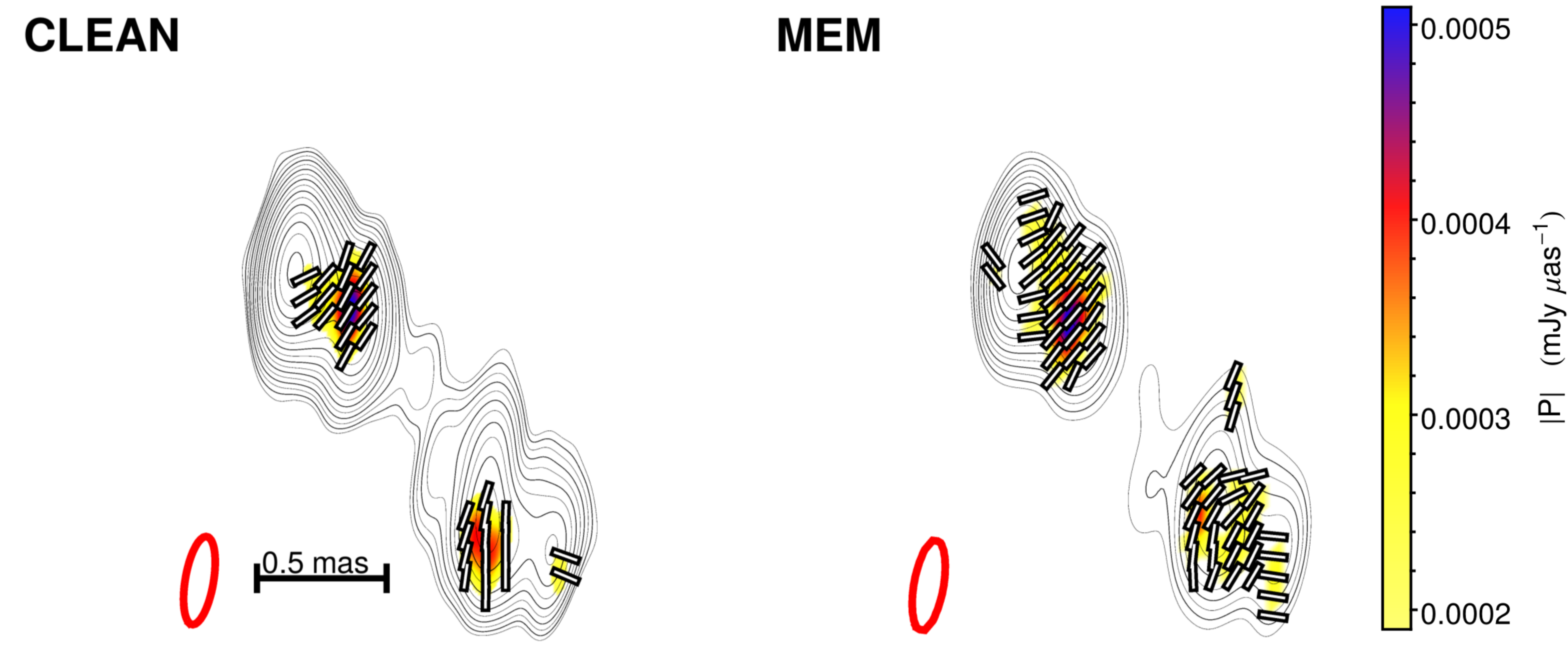}
\caption
{
Reconstructions of 3-mm observations of the quasar 3C273 taken with the VLBA+GBT \citep{Hada_2016}. Contours are of total flux in steps of $\sqrt{2}$ from 3$\times$ the background RMS level. Ticks representing the direction of the polarization position angle and color corresponding to the polarized intensity $|P|$ are plotted in regions where $|P|$ is greater than 4$\times$ its background RMS level. The left panel shows the reconstruction convolved with the fitted elliptical clean beam ($340 \times 108$ $\mu$as FWHM, produced with natural weighting), and the right panel displays a MEM reconstruction of the same data set, smoothed with the fitted clean beam. While the CLEAN reconstruction used a self-calibration loop to determine visibility phases on $\tilde{I}$, $\tilde{Q}$, and $\tilde{U}$, the MEM reconstruction directly used bispectrum and polarimetric ratio data. The MEM reconstruction used the Ponsonby-Nityananda-Narayan (PNN) entropy term.
}
\label{fig::3c273_86}
\end{figure*}

To further test our method at higher frequencies, we produced a polarimetric image from the 3~-mm observation of 3C273 taken with the VLBA in conjunction with the Green Bank Telescope (GBT) reported in \citet{Hada_2016}. The results are displayed in Fig.~\ref{fig::3c273_86}. When convolved with the clean beam, the results from our MEM method are broadly consistent, but some discrepancies are apparent in regions of low polarized intensity. 

\subsection{``Superresolution'' and Comparisons with CLEAN on Simulated 1.3-mm data}
\label{sec::cleancompare}

Unlike CLEAN images, MEM images in theory do not require restoration with the fitted interferometer beam. However, when testing our method on synthetic data, we found that at a certain point in the imaging process both total intensity and polarimetric MEM algorithms began producing spurious high-frequency features not present in the true source distribution. Restoring the final MEM images by convolving with a Gaussian beam will offset this tendency, but it is important not to make the beam too large and wash out real high-spatial-frequency features that may be ``superresolved'' on scales smaller than the interferometer beam.  

To test MEM's capacity for ``superresolution'' and determine the appropriate restoring beam size, we produced total intensity MEM (using the entropy term in Eq.~\ref{eq::Ssimple}) and CLEAN images of a model of \sgra\ using simulated data with thermal noise from the EHT array projected to be available in 2017 (See Section~\ref{sec::EHTobs}). For this simple test we neglected the effects of inaccurate amplitude calibration, atmospheric phase corruption, and interstellar scattering. We used a MEM algorithm with full visibility phase information, directly minimizing Eq~\ref{eq::objfunc} with the $\chi^2$ term in Eq.~\ref{eq::chi2}. This choice, while infeasible in practice due to phase errors, allowed us to directly compare to CLEAN without introducing the need for self-calibration.

After obtaining MEM and CLEAN reconstructions from the same data, we convolved the reconstructed images with a sequence of Gaussian beams scaled from the elliptical Gaussian fitted to the Fourier transform of the $u,v$ coverage (the ``clean beam'').   We then computed the normalized root-mean-square error (NRMSE) of each restored image:
\begin{align}
 \label{eq::mse}
 \text{NRMSE} = \sqrt{\frac{\sum_{i=1}^{n^2}|I'_i-I_i|^2}{\sum_{i=1}^{n^2}{|I_i|^2}}},
\end{align}
where $\mathbf{I'}$ is the final restored image and $\mathbf{I}$ is the true image. For the CLEAN reconstructions, we chose \emph{not} to add the dirty image residuals back to the convolved model, as the residuals are a sensible quantity only for the full restoring beam. To minimize the effect of this choice on the CLEAN reconstruction, we chose a compact model image with no diffuse structure. After tuning our CLEAN reconstruction for this image, the total flux left in the residuals was less than $2\%$ of the total image flux. In performing the CLEAN reconstruction, we used Briggs weighting and a loop gain of 0.025, with the rest of the parameters set to the default in the algorithm's CASA implementation \footnote{http://casa.nrao.edu/docs/TaskRef/clean-task.html}. 

\begin{figure*}[h]
\begin{centering}
\includegraphics*[width=\textwidth]{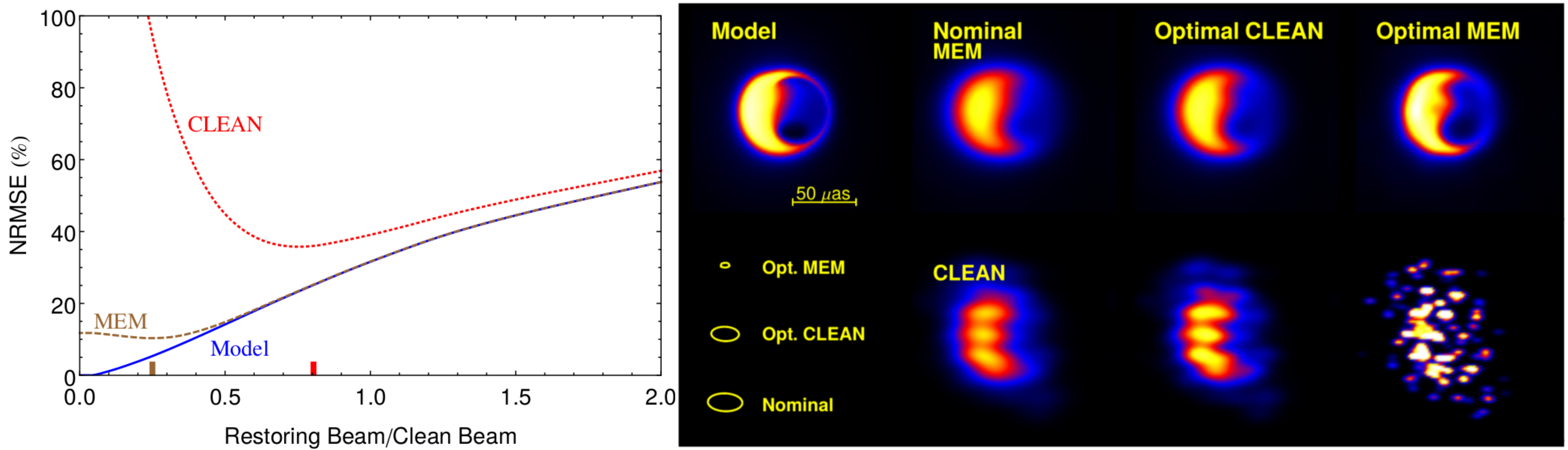}
\end{centering}
\caption{(Left) Normalized root-mean-square error (NRMSE, Eq.~\ref{eq::mse}) of MEM and CLEAN reconstructed Stokes $I$ images as a function of the fractional restoring beam size. For comparison, the NRMSE of the model image is also plotted. The reconstructed images were produced using simulated data from the EHT array; for straightforward comparison with CLEAN, realistic thermal noise was added to the simulated visibilities but gain calibration errors, random atmospheric phases, and blurring due to interstellar scattering were all neglected. The images were convolved with scaled versions of the fitted clean beam. The minimum for each NRMSE curve indicates the optimal restoring beam, which is significantly smaller for MEM (25\% of nominal) than for CLEAN (0.78\% of nominal). \\
(Right) Example reconstructions restored with scaled beams from curves in the left panel. The center-left panels are the MEM and CLEAN reconstructions restored at the nominal resolution, with the fitted clean beam. The center-right panels show the reconstructions restored with the optimal beam for the CLEAN reconstruction and the far right panels show both reconstructions restored with the optimal MEM beam. The CLEAN reconstructions consist of only the CLEAN components convolved with the restoring beam and do not include the dirty image residuals, as discussed at the end of Section~\ref{sec::IMEM}.
}
\label{fig::superres}
\end{figure*}

The results are displayed in Fig.~\ref{fig::superres}.  In the left panel, we see that the MEM curve has a minimum in NRMSE at a significantly smaller beam size than the CLEAN reconstruction, demonstrating MEM's superior ability to superresolve source structure over CLEAN. Furthermore, the value of NRMSE from the MEM reconstruction is consistently lower than from CLEAN for all values of restoring beam size. Most importantly, while the CLEAN curve NRMSE increases rapidly for restoring beams smaller than the optimal resolution, the MEM image fidelity is relatively unaffected by choosing a restoring beam that is too small. Choosing a restoring beam that is too large produces an image with the same fidelity as the model blurred to that resolution. The right panel of Fig.~\ref{fig::superres} shows the model image, the interferometer ``clean'' beam, and the reconstructions blurred with the clean beam (nominal) and the measured optimal fractional beams. In addition to lower resolution and fidelity, the CLEAN reconstructions show prominent striping features from isolated components being restored with the restoring beam. 

While Fig.~\ref{fig::superres} demonstrates that in this case the MEM reconstruction has superior resolution and fidelity to the CLEAN reconstruction, the optimal restoring beam size for the CLEAN reconstruction is still less than unity. This result was observed in several similar reconstructions, suggesting that shrinking the restoring beam used in CLEAN reconstructions to 75\% of the nominal fitted beam can enhance resolution without introducing imaging artifacts, at least on images of compact objects similar to those used in these tests. 

Repeating the exercise of Fig.~\ref{fig::superres} with observations taken with increased or decreased signal-to-noise ratio resulted in NRMSE curves that are only slightly higher and lower than the curves in Fig.~\ref{fig::superres}, but shared the same form - in particular, the location of the minimum NRMSE values was barely shifted. This insensitivity to additional noise is likely due to the overall high SNR of our original observations, which had an average SNR of 178 and a minimum SNR of 13. Our results show that with a high average SNR, increasing or decreasing the noise by up to an order of magnitude does not significantly affect the image reconstruction. Observations with an average SNR $\sim1$, on the other hand, may show a drastic change in quality with small adjustments to the noise level.

\begin{figure*}[h]
\begin{centering}
\includegraphics*[width=\textwidth]{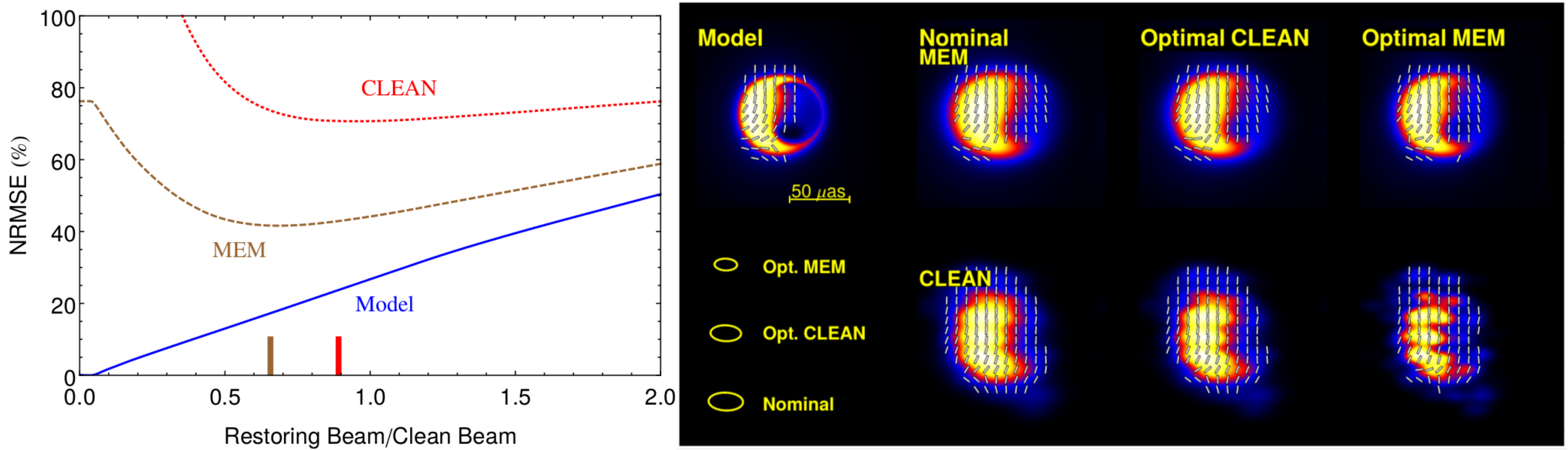}
\end{centering}
 \caption
{
(Left) Normalized root-mean-square error (Eq.~\ref{eq::mse}, with $I \rightarrow P$) of MEM and CLEAN reconstructed polarimetric images versus the size of the anisotropic restoring beam, as a fraction of the nominal fitted beam size. As in Fig.~\ref{fig::superres}, the CLEAN curve was computed by restoring the CLEAN point source model with scaled restoring beams without adding the dirty image residuals. The reconstructed images were produced using data simulated from the EHT array with realistic thermal noise; for simplicity of comparison with CLEAN, the data were not corrupted with gain uncertainties, random atmospheric phases, or blurring from interstellar scattering. Comparing to Fig.~\ref{fig::superres}, we see that by this metric the reconstruction of the linear polarization distribution  is less accurate than the reconstructions of Stokes $I$, but that the MEM reconstruction still provides superior resolution and fidelity to CLEAN, with optimal beam sizes at 70\% and 95\% of the nominal clean beam size, respectively.
(Right) Example reconstructions restored with scaled beams from curves in the left panel. The center-left panels are the MEM and CLEAN reconstructions restored at the nominal resolution, with the fitted clean beam. The center-right panels show the reconstructions restored with the optimal beam for the CLEAN reconstruction and the far right panels show both reconstructions restored with the optimal MEM beam. Polarization position angle ticks are plotted in regions with $I$ greater than 4$\times$  its RMS value and $|P|$ greater than 2$\times$ its RMS value. 
}
\label{fig::superrespol}
\end{figure*}

Extending the exercise from Fig.~\ref{fig::superres}, we calculated the NRMSE for polarimetric MEM from several test images as a function of restoring beam size, replacing the Stokes $I$ flux with $P = Q + iU$ in Eq.~\ref{eq::mse}. Once again, we neglected the effects of inaccurate amplitude calibration, atmospheric phase corruption, and interstellar scattering in our simulated data; however, our MEM algorithm used only polarimetric ratios $\breve{m}$ while CLEAN reconstructed $Q$ and $U$ separately with full $\tilde{Q}$ and $\tilde{U}$ amplitude and phase information. The results are displayed in Fig.~\ref{fig::superrespol}.  While not displayed, reconstructions using different regularizer terms (i.e. Eqs.~\ref{eq::HWent}, \ref{eq::Ssimplepol}, \ref{eq::tv}) performed similarly.  The degree of superresolution in the polarimetric MEM reconstructions is less than in the total intensity case, typically with a minimum in NRMSE around a restoring beam size of $1/2$ the nominal resolution. This reduced degree of superresolution is likely due to a combination of lower SNR on the polarized data points, the loss of absolute phase information in the MEM imaging process, and the low dynamic range of the $|m|$ images \citep{HoldThesis_1990}. 

The simple tests presented in Figs.~\ref{fig::superres} and \ref{fig::superrespol} suggest that MEM can ``superresolve'' source structure in $I$ and $P$ on scales greater than about $1/2$ the nominal interferometer resolution. Furthermore, they suggest that at least on this class of images, featuring compact flux distributions, MEM achieves superior resolution and fidelity to CLEAN. For the remainder of this work, we adopted a strategy of restoring both total intensity and polarimetric images with a scaled beam $1/2$ the size of the fitted beam.  

\subsection{Imaging Different Models of \sgra\ and M87 at 1.3-mm}
\label{sec::EHTobs}
We applied our techniques on several 1.3-mm simulated images from supermassive black hole accretion disk and jet models with simulated data from the planned 2017 EHT array. We chose several images featuring different types of structure in total intensity and polarization, including semi-analytic radiatively inefficient accretion flow (RIAF) and jet models courtesy of Avery Broderick \citep{Avery_Disk, Avery_Jet, Avery_Jet2}, ray-traced images from a magnetically arrested disk (see e.g. \citet{Tchek}) GRMHD simulation courtesy of Jason Dexter \citep{Dexter_2014}, and a GRMHD simulation from Roman Gold \citep{Gold_Pol, Sch_Sim}.

\begin{figure*}[h!]
 \centering
 \includegraphics*[width=0.75\textwidth]{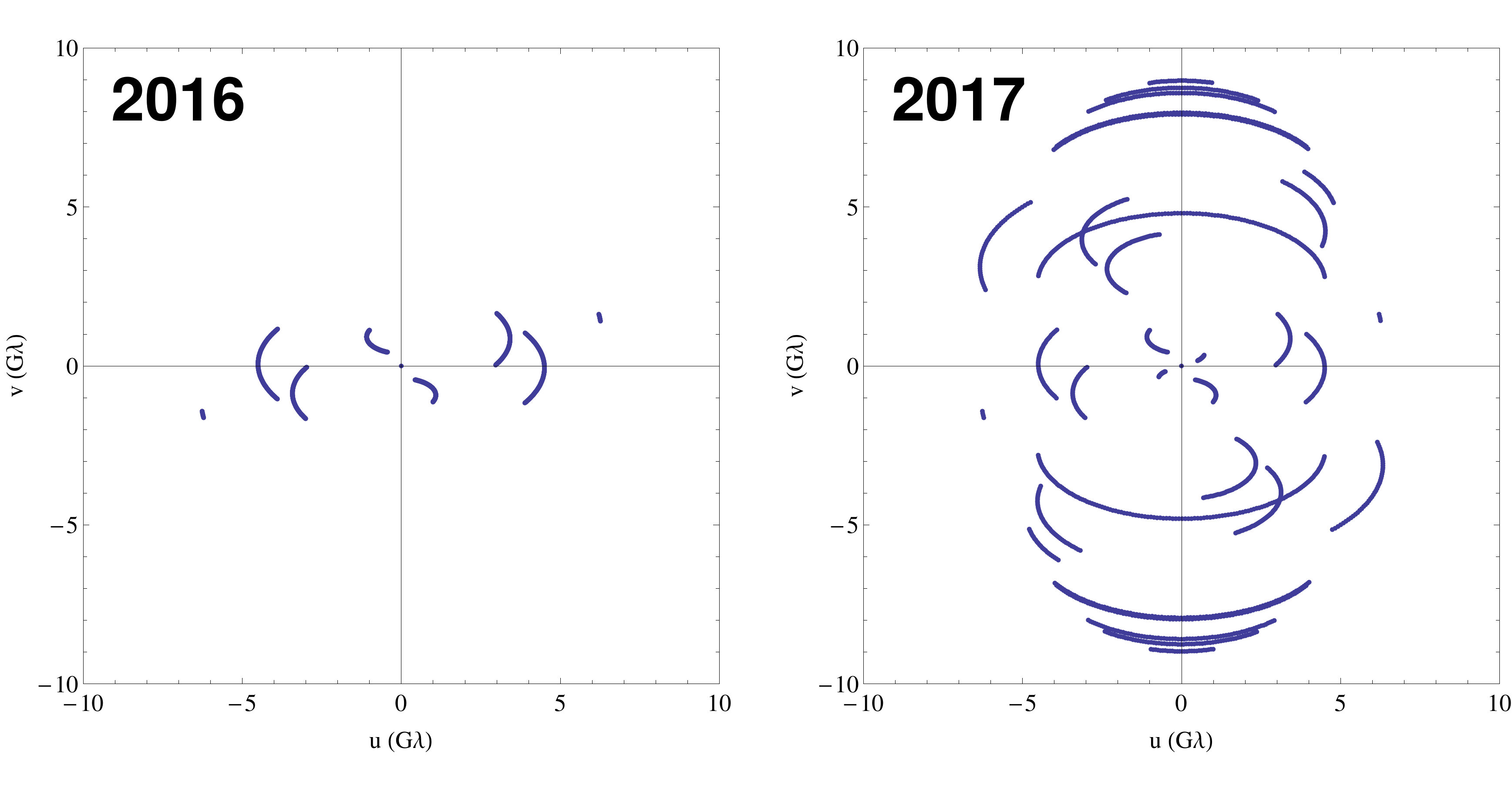}
 \caption
 {
 Event Horizon Telescope 24-hour $u,v$ coverage for observations of Sgr A* in 2016 (left) and 2017 (right). The 2016 array includes the Submillimeter Array in Hawaii, the Submillimeter Telescope in Arizona, the Large Millimeter Telescope in Mexico, and the Pico Veleta millimeter dish in Spain. In 2017 the array is projected to expand to include the Plateau de Bure interferometer in France, the ALMA interferometer in Chile, and the South Pole Telescope. 
 }
\label{fig::EHTuv_SgrA}
\end{figure*}

We sampled the Fourier transforms of these model images on projected baselines corresponding to the expected EHT arrays in 2016 and 2017. Our 2016 array included stations in Hawaii, Arizona, and Mexico, all operating with 2 GHz of bandwidth. The 2017 array is expected to include these stations with the addition of stations in France, the South Pole, and the ALMA interferometer in Chile, all operating with 4 GHz of bandwidth (See $u,v$ coverage in Fig.~\ref{fig::EHTuv_SgrA}). We added realistic baseline-dependent Gaussian thermal noise on the complex visibilities. The standard deviation $\sigma$ of the thermal noise is determined according to the standard equation \citepalias{TMS}
\begin{equation}
 \label{eq::noise}
 \sigma = \frac{1}{0.88} \sqrt{\frac{\text{SEFD}_1 \times\ \text{SEFD}_2}{2 \Delta \, \nu \, t_{\text{int}}}},
\end{equation}
where $\text{SEFD}_1$ and $\text{SEFD}_2$ are the telescope system equivalent flux densities, $\Delta \nu$ is the observing bandwidth, and $t_{\text{int}}$ is the integration time. The factor of $1/0.88$ in Eq.~\ref{eq::noise} comes from losses due to 2-bit quantization in the correlation process. For our simulations, we used an integration time of 60 s, unless otherwise stated.

\begin{figure*}[h]
 \centering
 \includegraphics*[width=\linewidth]{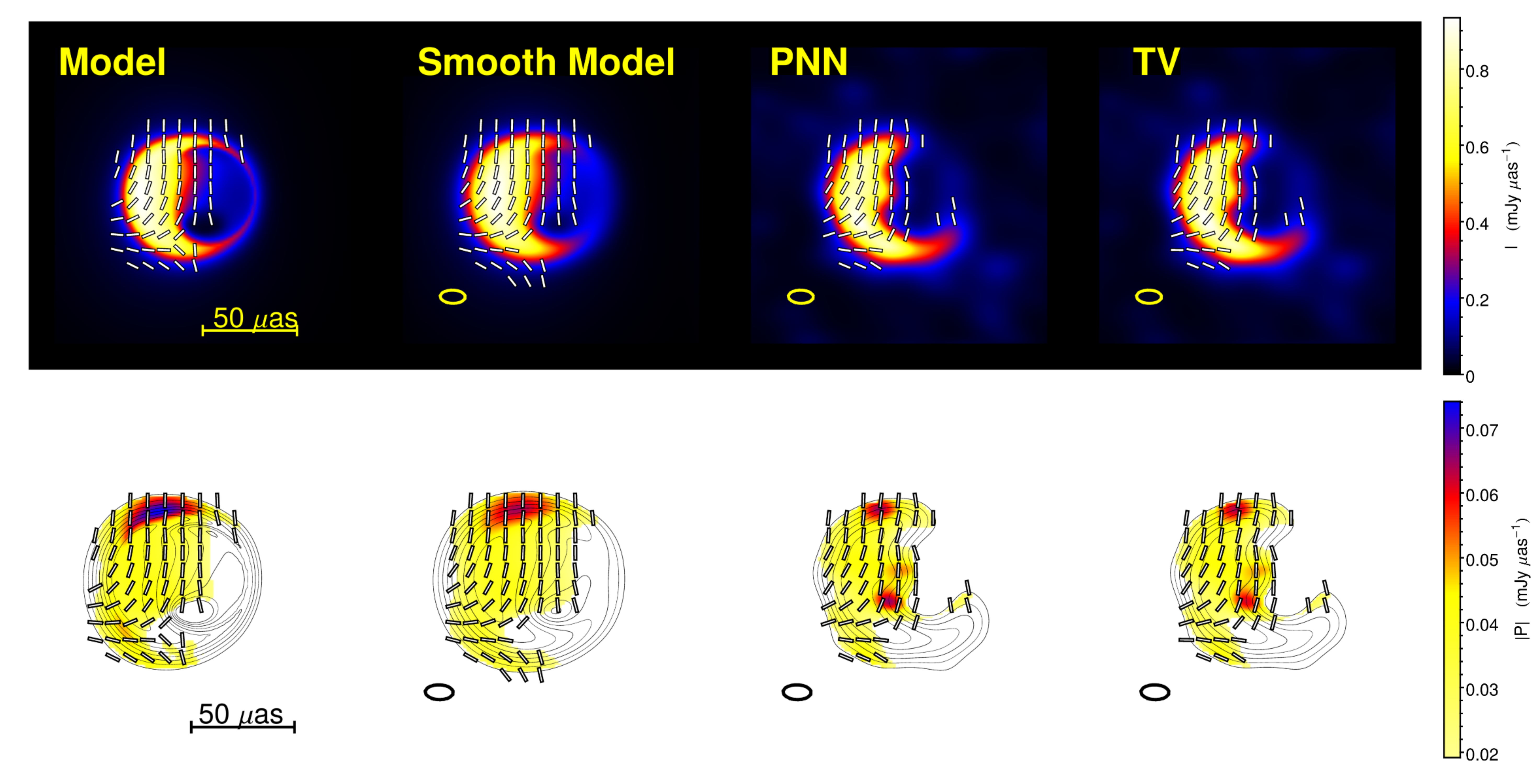}
 \caption
 {
 (Top) 1.3 mm MEM reconstructions of a \sgra\ image (left) from a simulation courtesy of Avery Broderick \citep{Avery_Disk}. Color indicates Stokes $I$ flux, and ticks marking the polarization position angle are plotted in regions with $I$ greater than 4$\times$ its RMS value and $|P|$ greater than 2$\times$ its RMS value. Visibilities from the planned full EHT array were simulated including the blurring effects of interstellar scattering, with realistic thermal noise, amplitude calibration errors, and random atmospheric phases included. Stokes $I$ was imaged  with the bispectrum and linear polarization was subsequently imaged using polarimetric ratios with the Ponsonby-Narayan-Nityananda (PNN) entropy function (right center) and a total variation (TV) regularizer (right). The final reconstructions were restored with a Gaussian beam 1/2 the size of the fitted clean beam ($27 \times 14$ $\mu$as FWHM); for comparison, the model image smoothed to this resolution is displayed on the center left. \\
 (Bottom) The same reconstructions displayed in contours of total intensity, in steps of $\sqrt{2}$ up from 4$\times$ the background RMS level. Color indicates the magnitude of the polarized flux $|P|$ and is displayed, along with ticks marking the polarization position angle, in regions where $I$ is greater than 4$\times$ its background RMS level and $|P|$ is greater than 2$\times$ its RMS value. Both MEM priors successfully reproduce the smooth polarization morphology of the simulated image.\\
 }
\label{fig::AveryImages}
\end{figure*}

We simulated the effects of gain calibration errors by assigning each site both a time dependent gain $G_i$ drawn from a Gaussian distribution with mean 1 and 10\% standard deviation and a time dependent atmospheric opacity $\tau_i$ drawn from a Gaussian with mean 0.1 and a standard deviation of 0.01. The ``true'' time-dependent SEFDs were then computed from the measured $\text{SEFD}'$s (denoted with primes) from the equation

\begin{equation}
 \label{eq::gain}
 \text{SEFD}_i = \text{SEFD}_i' \, \frac{e^{\tau_i / \sin{\theta_i}}}{G_i},
\end{equation}
where $\theta_i$ is the source elevation at the observation time. Thermal noise was added to the observations from a zero-mean circular complex Gaussian distribution of standard deviation given by Eq.~\ref{eq::noise}, with the measured $\text{SEFD}'$ replaced with the true SEFD at each time. The noisy visibilities were then multiplied by the ratio of the estimated to true SEFDs
\begin{equation}
 \label{eq::visgain}
 \tilde{I}_{ij} \rightarrow \tilde{I}_{ij} \times \sqrt{\frac{\text{SEFD}_i'e^{0.1 / \sin{\theta_i}}\,\text{SEFD}_j'e^{0.1 / \sin{\theta_i}}}{\text{SEFD}_i\,\text{SEFD}_j}},
\end{equation}
where we have used our assumption that the mean opacity at each site is 0.1 to adjust each of the estimated SEFDs for elevation dependence. The measured EHT station $\text{SEFD}'$s we used were reported in \citet{Lu_2014}. In computing the expected thermal noise with these $\text{SEFD}'$s by Eq.~\ref{eq::noise} for computing $\chi^2$ terms, we again modified each $\text{SEFD}'$ by the $e^{0.1 / \sin{\theta}}$ factor from elevation dependence, assuming an opacity $\tau = 0.1$. We did not include any correction for possible gain calibration error in our estimated noise terms.

In simulating phase corruption from atmospheric turbulence, we multiplied the visibilities by random phases drawn uniformly at each site and at each time step. For simulated observations of \sgra\ we also included the blurring effects of interstellar scattering, which we mitigated by dividing out the scattering kernel according to the method of \citet{Fish_2014}. This process has the net effect of increasing the noise level on long baselines. In practice, for Sgr A* refractive interstellar scattering contributes additional epoch-dependent image distortions \citep{Johnson_Scat}, which we will analyze separately in a future work. 

To compare the polarimetric reconstructions with different regularizers and with different arrays to the model image, we computed the NRMSE in Stokes $I$ and $P$ for each image via Eq.~\ref{eq::mse}. To compare the fidelity of the polarization position angle reconstruction, we also computed the mean square error of the polarization position angle of the reconstruction weighted by the magnitude of the total flux:
\begin{equation}
 \label{eq::angerr}
 \text{Weighted Angular Error} = \sqrt{\frac{\sum(\text{mod}(\chi'_i-\chi_i, \pi)^2\,|I_i|}{\sum|I_i|}}.
\end{equation}
This error metric gives an RMS estimate for the angular error in the polarization position angle reconstruction, and hence the magnetic field morphology of the source. It is weighted by the Stokes $I$ flux because the polarization position angle in the reconstructions can swing wildly in regions with negligible polarized flux. This reasoning also led us to display polarization position angle ticks only in pixels with greater than 10\% of the maximum Stokes $I$ flux in all of our reconstructed images.

\begin{figure*}[h]
 \centering
\includegraphics*[width=1\linewidth]{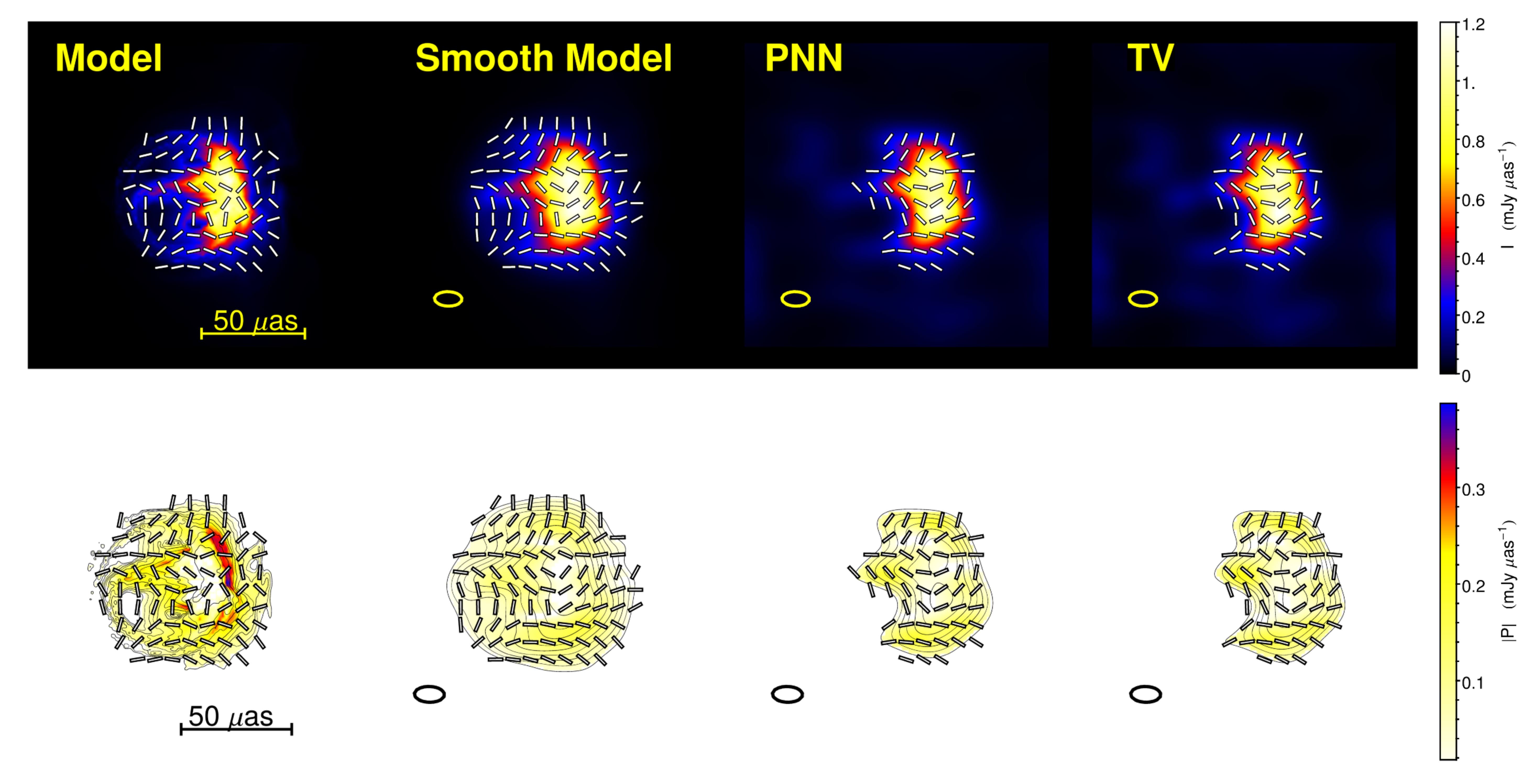}
 \caption
 {
 (Top) 1.3 mm MEM reconstructions of a ray-traced image computed from a GRMHD simulation of Sgr A* (left), provided courtesy of Roman Gold \citep{Gold_Pol}. Color indicates Stokes $I$ flux. Ticks marking the direction of linear polarization are displayed in regions with $I$ greater than 4$\times$ its RMS value and $|P|$ greater than 2$\times$ its RMS value. Visibilities from the planned full EHT array were simulated including the blurring effects of interstellar scattering, with realistic thermal noise, amplitude calibration errors, and random atmospheric phases included. Stokes $I$ was imaged  with the bispectrum and linear polarization was subsequently imaged using polarimetric ratios with the Ponsonby-Narayan-Nityananda (PNN) entropy function (right center) and a total variation (TV) regularizer (right). The final reconstructions were restored with a Gaussian beam 1/2 the size of the fitted clean beam ($27 \times 14$ $\mu$as FWHM); for comparison, the model image smoothed to this resolution is displayed on the center left.\\
 (Bottom) The same reconstructions displayed in contours of total intensity, in steps of $\sqrt{2}$ up from the 4$\times$ background RMS level. Color indicating the magnitude of the polarized flux, $|P|$, is displayed along with polarization position angle ticks in regions with $I$ greater than 4$\times$ its RMS value and $|P|$ greater than 2$\times$ its RMS value. The reconstructions more accurately reproduce the direction of linear polarization than the fractional polarization, as fractional polarization in the reconstructions tends to become large in regions of low total flux. Nonetheless, both reconstructions recover an accurate picture of global structure of the model polarized flux distribution blurred to the EHT's resolution.
 }
 \label{fig::RomanImages}
\end{figure*}

The EHT simulated data has lower signal-to-noise than the VLBA data considered above, but we found that our MEM reconstructions were still nearly independent of the choice of regularizer and relative weighting. The polarimetric reconstructions were able to conclusively distinguish between the well-ordered field structure in a RIAF model (Fig.~\ref{fig::AveryImages}) and the stochastic field configuration in a GRMHD simulation (Fig.~\ref{fig::RomanImages}). Both the PNN and TV entropy terms reproduce the polarization magnitude and direction well, and the NRMSE in both $I$ and $P$ (Eq.~\ref{eq::mse}) and the intensity-weighted polarization position angle error (Eq.~\ref{eq::angerr}) were similar for reconstructions with both entropy terms. For the RIAF model, which featured low polarization magnitudes and smoothly varying polarization position angle, the NRMSE values were 28.7\% for Stokes $I$ and 47.9\% for Stokes $P$ for the PNN reconstruction and 28.53\% in Stokes $I$ and 48.2\% for $P$ for the TV reconstruction. The corresponding weighted angular errors were $14.9^\circ$ and $14.7^\circ$.  When we compared the reconstructions to the model image smoothed to the same resolution as the reconstruction's resolution (center left in Fig.~\ref{fig::AveryImages}), the $I$ and $P$ NRMSE values drop to 25.1\% and 46.7\% for the PNN reconstruction and 24.8\% and 47.0\% for the TV reconstruction. The polarization position angle weighted error drops to $13.7^\circ$ and $13.5^\circ$ for the PNN and TV regularizers, respectively.

For the disordered field in the GRMHD simulation (Fig.~\ref{fig::RomanImages}), the NRMSE and weighted angular error fidelity metrics again give similar results for both reconstructions, but slightly favor the PNN image. For PNN, the NRMSE values were 30.4\% for Stokes $I$ and 74.23\% for Stokes $P$, with a weighted angular error of $34.1^\circ$. For TV, the NRMSE values were 32.0\% for Stokes $I$ and 76.0\% for $P$, with a weighted angular error of $34.3^\circ$. These  high angular error values occur because of the mismatch of the smoothed-out polarization field of the reconstruction and the fine-scale structure in the model image. When compared to the smoothed model image at (center left in Fig.~\ref{fig::RomanImages}, the Stokes $I$ and $P$ NRMSE drop to 22.6\% and 39.9\% for the PNN reconstruction and 24.8\% and 46.0\% for the TV reconstruction; the polarization position angle weighted errors drop to $17.2^\circ$ for the PNN and $18.5^\circ$ for TV. 

\begin{figure*}[h]
 \centering
\includegraphics*[width=1\linewidth]{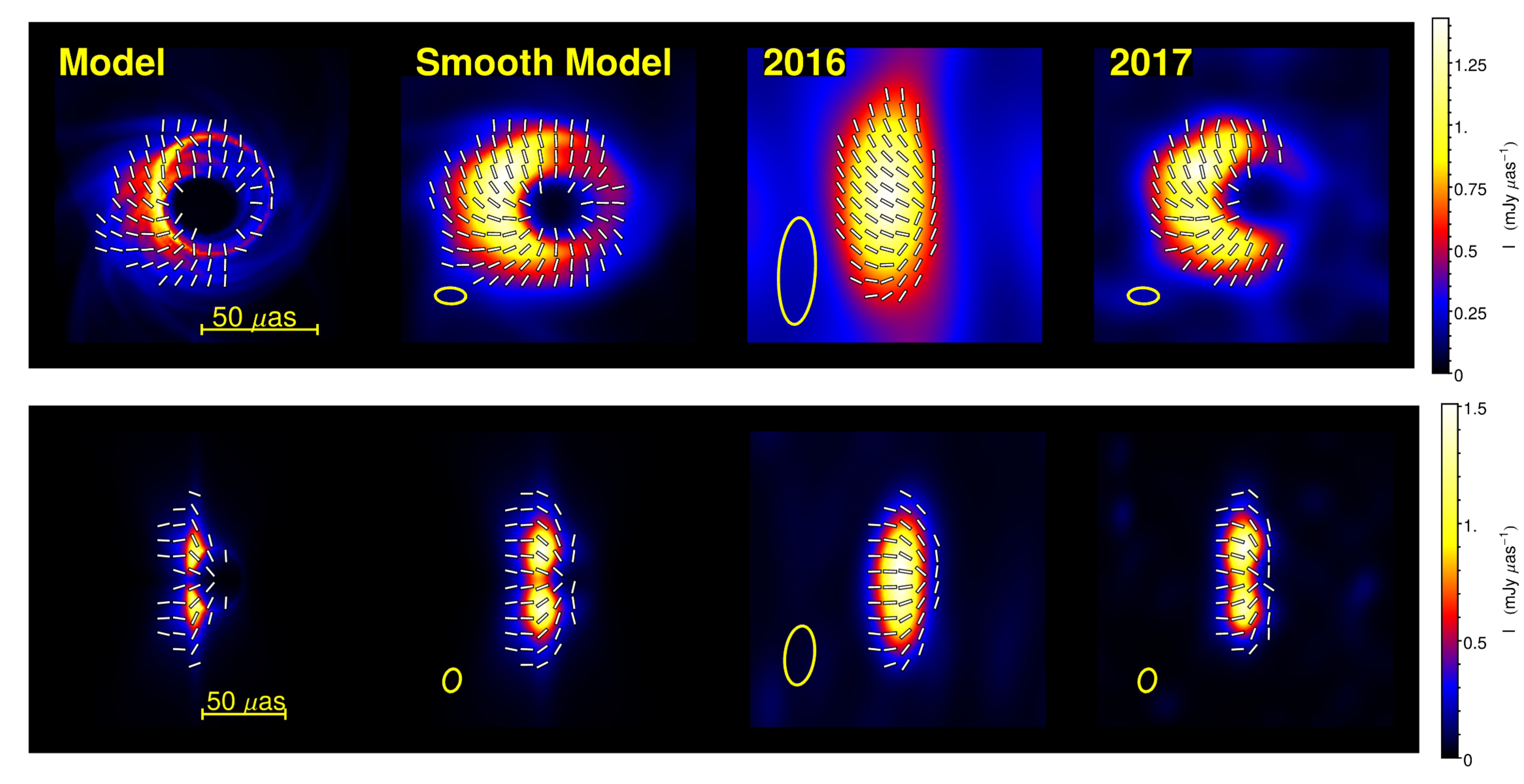}
 \caption
 {
 (Top) 1.3-mm MEM reconstructions of a magnetically arrested disk simulation of the Sgr A* accretion flow, courtesy of Jason Dexter \citep{Dexter_2014}. Color indicates Stokes $I$ flux and ticks marking the direction of linear polarization are plotted in regions with $I$ greater than 4$\times$ its RMS value and $|P|$ greater than 2$\times$ its RMS value. After blurring the image with the Sgr A* scattering kernel at 1.3 mm, data were simulated with realistic thermal noise, amplitude calibration errors, and random atmospheric phases. The center right panel shows a reconstruction with data simulated on EHT baselines expected in 2016 and the rightmost panel shows the reconstruction with the full array expected in 2017. Each reconstruction was restored with a Gaussian beam 1/2 the size of the fitted clean beam ($93 \times 32$ $\mu$as FWHM in 2016 ; $27 \times 14$ $\mu$as FWHM in 2017). For comparison, the center left panel shows the model smoothed to the same resolution as the 2017 image.  
 (Bottom) 1.3-mm MEM reconstructions of a simulation of the jet in M87, courtesy of Avery Broderick \citep{Avery_Jet, Avery_Jet2}. Data were simulated on 2016 and 2017 EHT baselines as in the top panel, but without the contributions from interstellar scattering that are significant for \sgra. \ Both reconstructions were restored with a Gaussian beam 1/2 the size of the fitted clean beam ($72 \times 36$ $\mu$as FWHM in 2016 ; $28 \times 20$ $\mu$as FWHM in 2017).
 }
\label{fig::ArrayCompare}
\end{figure*}

We also compared images produced with the PNN regularizer using simulated data from the full 2017 array and the smaller four-element array that observed in March of 2016 (Fig.\ref{fig::EHTuv_SgrA}, left panel). As expected, the fidelity metrics show distinct improvement between the 2016 and 2017 reconstructions. Both in simulations of the near-horizon jet emission in M87 and accretion disk emission in \sgra\ (Fig.~\ref{fig::ArrayCompare}), we found that the larger amount of information in polarimetric VLBI data over total intensity visibilities (due to the ability to accurately calibrate the phases of polarimetric ratios) was significant with the sparse baseline coverage in 2016. Namely, we were able to achieve more detail in the polarized emission reconstruction than its total intensity counterpart in both cases. In the absence of long baselines needed to resolve distinguishing features in the total intensity image, polarimetric imaging can help distinguish between different models of the emission region such as the disk and jet models in Fig.~\ref{fig::ArrayCompare}. Polarimetric images, even with poor resolution, can begin to characterize the general magnetic field structure in \sgra\ and M87 with near-term EHT observations even before completely resolving the emission region or black hole shadow.

For our test \sgra\ model, a magnetically arrested disk GRMHD simulation (Fig.~\ref{fig::ArrayCompare}, top panel), the NRMSE of the reconstructions shows distinct improvement between 2016 and 2017, but the weighted angular error (Eq.~\ref{eq::angerr}) metric is surprisingly similar across the reconstructions. For 2016, the NRMSE values were 52.30\% for Stokes $I$ and 77.3\% for Stokes $P$, with a weighted angular error of $29.3^\circ$. In 2017, the NRMSE values were 36.06\% for Stokes $I$ and 66.9\% for $P$, with an angular error of $28.3^\circ$. While the 2017 array long, high-sensitivity baselines to the ALMA array produces a qualitatively and quantitatively superior reconstruction, MEM techniques reproduce qualitative features of the polarization structure even with the sparse 2016 data. 

When we instead compare the reconstructions to the model image smoothed to the same resolution as the respective restoring beam, the $I$ and $P$ NRMSE values drop to 24.0\% and 59.0\% for the 2016 reconstruction and 19.8\% and 61.9\% for the 2017 image. The polarization position angle weighted error drops to $20.0^\circ$ and $21.6^\circ$ for the 2016 and 2017 images, respectively. Even with minimal baseline coverage, MEM is able to reconstruct a reasonably accurate image when compared to the true image viewed at the same resolution. 

The 2016 image of an M87 jet model (Fig.~\ref{fig::ArrayCompare}, bottom panel) gave NRMSE values of 55.61\% for Stokes $I$ and 77.34\% for Stokes $P$, with a weighted angular error of $23.5^\circ$. In 2017, the NRMSE values were 36.71\% for Stokes $I$ and 54.40\% for $P$, with an angular error of $17.9^\circ$.  When we instead compare the reconstructions to the model image smoothed to the same resolution as the restoring beam, the $I$ and $P$ NRMSE values drop to 21.3\% and 34.5\%  for the 2016 image and 18.3\% and 27.7\% for the 2017 image, while the polarization position angle weighted error drops to $21.6^\circ$ and $14.8^\circ$ for the 2016 and 2017 images, respectively. 

\section{Conclusion}
\label{sec::Summary}
As the EHT opens up new, extreme environments to direct VLBI imaging, a renewed exploration of VLBI imaging strategies is necessary for extracting physical signatures from challenging datasets. In this paper, we have shown the effectiveness of imaging linear polarization from VLBI data using extensions of the Maximum Entropy Method. We explored extensions of MEM using previously proposed polarimetric regularizers like PNN and adaptations of regularizers new to VLBI imaging like total variation. We furthermore adapted standard MEM to operate on robust bispectrum and polarimetric ratio measurements instead of calibrated visibilities. MEM imaging of polarization can provide increased resolution over CLEAN (Fig.~\ref{fig::superrespol}) and is more adapted to continuous distributions, as are expected for the black hole accretion disks and jets targeted by the Event Horizon Telescope. Furthermore, MEM imaging algorithms can naturally incorporate both physical constraints on flux and polarization fraction as well as constraints from prior information or expected source structure. Extending our code to run on data from connected-element interferometers like ALMA is a logical next step, but it will require new methods to efficiently handle large amount of data and image pixels across a wide field of view. Polarimetric MEM is also a promising tool for synthesis imaging of a diversity of other astrophysical systems typically observed with connected element interferometers. For example, the polarized dust emission from protostellar cores frequently exhibits a smooth morphology \citep{Girart_2006, Hull_2013}, so MEM may be better-suited to study both the large-scale magnetic-field morphologies and their small deviations than typical reconstructions using CLEAN.

The natural ability to incorporate various image constraints makes extensions of MEM useful for investigating new forms of image reconstruction that will be relevant for future EHT observations. Although our algorithm is relatively insensitive to calibration errors and we have shown that our reconstructions are reliable even after including realistic station gain uncertainties and fluctuations, we have not yet incorporated amplitude self-calibration that could further improve reconstructions of $\tilde{I}$. Future work will also investigate a Stokes $I$ MEM imaging algorithm that uses only closure phase and closure amplitude data, which would be immune to phase and amplitude calibration errors, thereby eliminating the need for self calibration.  Another goal is addition of dynamic deblurring that can disentangle effects of strong interstellar scattering with more complicated structure than the simple convolution that holds in the long-term average regime \citep{Johnson_Scat}. With polarization, MEM could be used to image Faraday rotation across a frequency band. As an imaging framework, MEM is highly flexible and we expect that continued investigation will lead to new algorithms that can be tailored to the particular challenges of EHT image reconstruction.

\acknowledgments

We thank the National Science Foundation (AST-1310896, AST-1312034, AST-1211539, and AST-1440254) and the Gordon and Betty Moore Foundation (\#GBMF-3561) for financial support of this work. RN's research was supported in part by NSF grant AST1312651 and NASA grant TCAN NNX14AB47G. KB was supported by NSF CGV-1111415 and a NSF Graduate Fellowship. We thank  Svetlana Jorstad, Alan Marscher, and Kazuhiro Hada for providing the data imaged in Section~\ref{sec::bu} and for their helpful comments. We thank Avery Broderick, Jason Dexter, and  Roman Gold for generously providing model images. We also thank Lindy Blackburn for his help on simulating gain and phase errors and Kazunori Akiyama for his suggestion of applying total variation as a polarimetric regularizer. We thank the anonymous referee, whose thorough suggestions significantly improved this paper. This study makes use of 43 GHz VLBA data from the VLBA-BU Blazar Monitoring Program (VLBA-BU-BLAZAR; http://www.bu.edu/blazars/VLBAproject.html), funded by NASA through the Fermi Guest Investigator Program. The VLBA is an instrument of the National Radio Astronomy Observatory. The National Radio Astronomy Observatory is a facility of the National Science Foundation operated by Associated Universities, Inc.

\clearpage
\appendix
\section{Polarimetric VLBI observables}
In practice, visibilities are estimated by correlating the measured electric fields at different sites. In VLBI, circular feeds are most common, and the total-intensity visibility $\tilde{I}$ is then given as the average of the parallel-hand correlations while $\tilde{Q}_k$ and $\tilde{U}_k$ are estimated using combinations of the cross-hand visibilities. In terms of the cross-hand correlations at sites 1 and 2, the four interferometric Stokes parameters measured on the 1-2 baseline are \citep{RWB_1994}
\begin{align}
 \label{IQUV}
 \tilde{I}_{12} &= \frac{1}{2}\left(\left<R_1R_2^*\right> + \left<L_1L_2^*\right>\right) \\
 \tilde{Q}_{12} &= \frac{1}{2}\left(\left<L_1R_2^*\right> + \left<R_1L_2^*\right>\right) \\
 \tilde{U}_{12} &= \frac{i}{2}\left(\left<L_1R_2^*\right> - \left<R_1L_2^*\right>\right) \\
 \tilde{V}_{12} &= \frac{1}{2}\left(\left<R_1R_2^*\right> - \left<L_1L_2^*\right>\right). 
\end{align}
We ignore circular polarization in what follows. By the van Cittert-Zernike theorem (Eq.~\ref{eq::VCZ}), the complex visibilities $\tilde{I}(u,v), \tilde{Q}(u,v), \tilde{U}(u,v)$ are the Fourier transforms of the separate Stokes images $I(x,y), Q(x,y), U(x,y)$. 
The image linear polarization can also be represented with the fractional polarization $m$ and polarization position angle $\chi$ (conventionally measured East of North)
where
\begin{equation}
 \label{mchi}
 m(x,y) = \frac{\sqrt{Q(x,y)^2+U(x,y)^2}}{I(x,y)} \; , \; \chi(x,y) = \frac{1}{2}\arctan{\frac{U(x,y)}{Q(x,y)}}.
\end{equation}
The distinction between the polarization position angle $\chi$ and the data term $\chi^2$ should be clear from the context. Similarly, we can decompose the Fourier conjugate $\tilde{P}$ (Eq.~\ref{eq::polrat})
\begin{equation}
 \label{Pdef}
 \tilde{P}(u,v) = \tilde{Q}(u,v) + i\tilde{U}(u,v) = \tilde{I}(u,v)\breve{m}(u,v).
\end{equation}
Again, note that the complex quantity $\breve{m}(u,v)$ is not the Fourier conjugate of the real position-space fractional polarization $m(x,y)$.

Since $I,Q,U$ are real, $\tilde{I},\tilde{Q},\tilde{U}$ are conjugate-symmetric under $(u,v)\rightarrow (-u,-v)$. This is not the case for $\tilde{P} = \tilde{Q} + i\tilde{U}$. Instead, using Eq.~\ref{IQUV}, for telescopes 1,2 corresponding to a baseline vector $(u,v)$, we see that
\begin{align}
 \label{Pconj}
 \tilde{P}(u,v)&=\left<R_1L_2^*\right> \\ 
 \tilde{P}(-u,-v)&=\left<L_1^*R_2\right>. 
\end{align}

In an imaging algorithm, we model the $n\times n$ total intensity and polarization images with length $n^2$ arrays $\mathbf{I'}$, $\mathbf{P'}$. The $N$ measured total intensity and polarimetric visibilities form arrays $\tilde{\mathbf{I}}$, $\tilde{\mathbf{P}}$. When comparing our measurements to a test image, we use the arrays of the sample visibilities $\mathbf{\tilde{I}'} = \mathbf{A}\mathbf{I'}$ and $\mathbf{\tilde{P}'} = \mathbf{A}\mathbf{P'}$, where $\mathbf{A}$ is a Fourier matrix
\begin{align}
 A_{ij}&=e^{-2\pi i (u_ix_j + v_iy_j)}.
\end{align}

\section{Thermal Noise}
Thermal noise on a VLBI baseline produces circular Gaussian error in the visibility plane with standard deviation $\sigma$ given by Eq.~\ref{eq::noise}. In principle, the thermal noise $\sigma$ is the same for $\tilde{I}$, $\tilde{Q}$, and $\tilde{U}$. The factor of $1/0.88$ comes from losses due to 2-bit quantization \citepalias{TMS}. The error in $\tilde{P}$ is also circular, with standard deviation:
\begin{equation}
 \label{perr}
 \sigma_P=\sqrt{2}\sigma.
\end{equation}
Since the error is assumed to be circular in the high SNR limit, the error in the visibility amplitude $|\tilde{I}|$ is equal to the error in the real and imaginary parts, and the error in the visibility phase $\phi$ is
\begin{equation}
 \label{erramp,phase}
 \delta|\tilde{I}|=\sigma \;,\;\delta\phi=\frac{\sigma}{|\tilde{I}|}
\end{equation}

The thermal noise on the bispectrum (Eq.~\ref{eq::bispec}) will in general not be described by a circular Gaussian distribution. However, in the limit of high SNR, we can approximate the distribution as a circular Gaussian with standard deviation \citepalias{TMS}
\begin{equation}
 \label{bierr}
 \sigma_B=\delta|\tilde{I}_B| =  |\tilde{I}_B|\sqrt{\frac{\sigma_1^2}{|\tilde{I}_1|^2}+\frac{\sigma_2^2}{|\tilde{I}_2|^2}+\frac{\sigma_3^2}{|\tilde{I}_3|^2}},
\end{equation}
where $|\tilde{I}_1|$, $|\tilde{I}_2|$, $|\tilde{I}_3|$, are the visibility amplitudes on the three baselines that make up the bispectrum, $\sigma_1$, $\sigma_2$, and $\sigma_3$ are their corresponding standard deviations, and $|\tilde{I}_B|=|\tilde{I}_1||\tilde{I}_2||\tilde{I}_3|$ is the bispectral amplitude. The error in the closure phase is just 
\begin{equation}
 \sigma_{c} = \frac{\sigma_B}{|\tilde{I}_B|}.
\end{equation}

Similarly, the distribution of the visibility domain polarimetric ratio $\tilde{P}/\tilde{I} = \breve{m}$ is not generally a complex circular Gaussian. In the limit of high SNR, however, we again approximate it as such with standard deviation given by
\begin{align}
 \label{mbreveerror}
 \sigma_m &= \delta|\breve{m}| = \sigma\sqrt{\frac{2}{|\tilde{I}|^2}+\frac{|\tilde{P}|^2}{|\tilde{I}|^4}}.
\end{align}

\section{Entropy and $\chi^2$ Gradients}
\label{sec::grad}
Our implementation of the maximum entropy method uses a gradient descent algorithm to minimize the objective function $J$. For total intensity imaging, the necessary gradients of $\chi^2_B$ with respect to the image pixels $I_k$ can be found in \citet{Katie_2015}. Below we list the gradients of the polarimetric $\chi^2$ and entropy terms used in the polarimetric imaging step.
 
\subsection{Polarimetric Ratio $\chi^2$}
In imaging $P$, we work directly with interferometric polarimetric ratios. The reduced $\chi_m^2(\mathbf{I'},\mathbf{P'})$ we use is (Eq.~\ref{eq::chi2m})
\begin{align}
 \label{rchisqm}
 \chi^2(\mathbf{I'},\mathbf{P'}) &=\frac{1}{2N}\sum_i^N\frac{|\tilde{P}_i/\tilde{I}_i-\tilde{P}'_i/\tilde{I}'_i|^2}{\sigma^2_i} = \frac{1}{2N}\sum_i^N\frac{|\breve{m}_i-\breve{m}'_i|^2}{\sigma^2_i},
\end{align}
where errors on the polarimetric ratios are calculated according to Eq.~\ref{mbreveerror}. Computing the gradient with respect to the image domain fractional polarizations $m_k$ and polarization position angles $\chi_k$ gives
\begin{align}
 \label{rchisqmgradm}
 \frac{\partial}{\partial m_k}\chi^2 &=-\frac{1}{N}\operatorname{Re}\left[I_ke^{-2i\chi_k}\sum_{i}A^\dagger_{ki}\left(\frac{\tilde{P}_i/\tilde{I}_i-\tilde{P}'_i/\tilde{I}'_i}{\tilde{I}^{'*}_i \, \sigma^2_i}\right)\right]
\end{align}

\begin{align}
 \label{rchisqmgradchi}
 \frac{\partial}{\partial \chi_k}\chi^2 &=-\frac{2}{N}\operatorname{Im}\left[I_km_ke^{-2i\chi_k}\sum_{i}A^\dagger_{ki}\left(\frac{\tilde{P}_i/\tilde{I}_i-\tilde{P}'_i/\tilde{I}'_i}{\tilde{I}^{'*}_i \, \sigma^2_i}\right)\right].
\end{align}


\subsection{Ponsonby/Nityananda/Narayan Entropy}
Setting $m_{\text{max}} = 1$ in Eq.~\ref{eq::HWent}, we have the traditional form of the PNN entropy:
\begin{align}
 \label{HWent}
 S(\mathbf{I'},\mathbf{P'}) = -\sum I'_i  \left( \frac{1+m'_i}{2}\log\frac{1+m'_i}{2} + \frac{1-m'_i}{2}\log\frac{1-m'_i}{2} \right).
\end{align}
It has gradients with respect to polarimetric ratio $m$ and polarization position angle $\chi$ given by
\begin{align}
 \label{HWgrad}
 \frac{\partial S}{\partial m'_k} &=-I'_k\operatorname{arctanh} m'_k \\
 \frac{\partial S}{\partial \chi'_k} &= 0.
\end{align}

\subsection{Total Variation Entropy}
Because Total Variation (Eq.~\ref{eq::tv}) involves differences between pixels in both the $x$ and $y$ image direction, we must adjust our notation to account for the 2D nature of the image. With both dimensions restored, the complex polarized image is $P_{i,j} = I_{i,j}\,m_{i,j}\,e^{2i\chi_{i,j}}$. The total variation entropy term is then
\begin{equation}
 \label{Stv}
 S(\mathbf{I'},\mathbf{P'})=-\sum_{ij}\sqrt{|P'_{i+1,j}-P'_{i,j}|^2+|P'_{i,j+1}-P'_{i,j}|^2}.
\end{equation}
The gradients with respect to $m$ and $\chi$ are
\begin{align}
 \label{Stvgrad}
 \frac{\partial S}{\partial m'_{k,l}} = &-\frac{2|P'_{k,l}|-|P'_{k+1,l}|\cos[2(\chi'_{k+1,l}-\chi'_{k,l})]-|P'_{k,l+1}|\cos[2(\chi'_{k,l+1}-\chi'_{k,l})]}{\sqrt{|P'_{k+1,l}-P'_{k,l}|^2 + |P'_{k,l+1}-P'_{k,l}|^2}} \nonumber\\
 &- \frac{|P'_{k,l}|-|P'_{k-1,l}|\cos[2(\chi'_{k,l}-\chi'_{k-1,l})]}{\sqrt{|P'_{k,l}-P'_{k-1,l}|^2 + |P'_{k-1,l}-P'_{k-1,l+1}|^2}} \nonumber\\
 &- \frac{|'P_{k,l}|-|P'_{k,l-1}|\cos[2(\chi'_{k,l}-\chi'_{k,l-1})]}{ \sqrt{|P'_{k+1,l-1}-P'_{k,l-1}|^2 + |P'_{k,l}-P'_{k,l-1}|^2}} \\ \nonumber \\
 \frac{\partial S}{\partial \chi'_{k,l}} = &+\frac{2|P'_{k,l}P'_{k+1,l}|\sin[2(\chi'_{k+1,l}-\chi'_{k,l})]+2|P'_{k,l}P'_{k,l+1}|\sin[2(\chi'_{k,l+1}-\chi'_{k,l})]}{\sqrt{|P'_{k+1,l}-P'_{k,l}|^2 + |P'_{k,l+1}-P'_{k,l}|^2}} \nonumber\\
 &- \frac{2|P'_{k,l}P'_{k-1,l}|\sin[2(\chi'_{k,l}-\chi'_{k-1,l})]}{\sqrt{|P'_{k,l}-P'_{k-1,l}|^2 + |P'_{k-1,l}-P'_{k-1,l+1}|^2}} \nonumber\\
 &- \frac{2|P'_{k,l}P'_{k,l-1}|\sin[2(\chi'_{k,l}-\chi'_{k,l-1})]}{\sqrt{|P'_{k+1,l-1}-P'_{k,l-1}|^2 + |P'_{k,l}-P'_{k,l-1}|^2}} 
\end{align}

\section{Image Change of Variables}
\label{sec::cov}
To extend the range of the image variables for total intensity and polarization to the entire real line and avoid the use of bounded minimization, we use the change of variables
\begin{equation}
\label{eq::cov}
 I_i = e^{\xi_i} \;\;,\;\; m_i = \frac{1}{2} + \frac{1}{\pi}\tan^{-1}\kappa_i.
\end{equation}
Consequently, we need to multiply the gradients given in section~\ref{sec::grad} by the chain rule factors
\begin{equation}
 \frac{d I_i}{d \xi_i} = e^{\xi_i} \;\;,\;\; \frac{d m_i}{d \kappa_i} = \frac{1}{\pi(1+\kappa_i^2)}.
\end{equation}
\newpage
\bibliography{PolMEM.bib}

\end{document}